%% file: main.tex
\documentclass[twocolumn]{aastex7}

\usepackage{amsmath,color,textcomp,url,graphicx,subfigure}
\usepackage{listings}
\usepackage{booktabs}
\usepackage{ifpdf}

    \newcommand{\nhqspectra}{6003}

    \newcommand{\nhqcand}{3953}

    \newcommand{\nhqknown}{2050}

    \newcommand{\nbrandnewcand}{945}

    \newcommand{\ncanducd}{1916}

    \newcommand{\nphotobd}{3008}

    \newcommand{\nbyw}{1937}

    \newcommand{\nspecial}{3739}

    \newcommand{\nknyoung}{125}

    \newcommand{\nnewyoung}{2668}

    \newcommand{\nknsd}{25}

    \newcommand{\nnewsd}{250}

    \newcommand{\nknpec}{10}

    \newcommand{\nnewpec}{865}

    \newcommand{\nspecucdbefore}{3449}

    \newcommand{\nspecucdafter}{7402}

\newcommand{\wmagsnr}{14.0}
\newcommand{\datapointslimit}{50}
\newcommand{\autotypemad}{1.5}
\newcommand{\percfailurestar}{8.6}
\newcommand{\percfailurescatter}{2.6}
\newcommand{\percfailuresnr}{0.3}
\newcommand{\percfailurereddened}{0.3}
\newcommand{\percfailbdplusstar}{1.1}

\newcommand{\completiontimestamp}{2026 March 11}

\newcommand{\percgood}{81.2}
\newcommand{\percbad}{7.6}
\newcommand{\percamb}{1.5}

\newcommand{\ntemplates}{47}
\newcommand{\chilimit}{3}
\newcommand{\smallpenalty}{12.4}
\newcommand{\largepenalty}{26.2}

\newcommand{\masyr}{$\mathrm{mas}\,\mathrm{yr}^{-1}$}

\newcommand{\planetarium}{Plan\'etarium de Montr\'eal, Espace pour la Vie, 4801 av. Pierre-de Coubertin, Montr\'eal, Qu\'ebec, Canada}
\newcommand{\irex}{Trottier Institute for Research on Exoplanets, Universit\'e de Montr\'eal, D\'epartement de Physique, C.P.~6128 Succ. Centre-ville, Montr\'eal, QC H3C~3J7, Canada}
\newcommand{\amnh}{Department of Astrophysics, American Museum of Natural History, Central Park West, New York, NY, USA}
\newcommand{\openu}{School of Physical Sciences, The Open University, Milton Keynes, MK7 6AA, UK}

\received{2026 March 18}
\revised{2026 April 23}

\submitjournal{ApJ}

\shorttitle{A SPHEREx Pipeline and Spectral Library for Ultracool Dwarfs}
\shortauthors{Gagn\'e et al.}

\begin{document}

\title{A SPHEREX PIPELINE AND SPECTRAL LIBRARY FOR ULTRACOOL DWARFS}

\author[0000-0002-2592-9612]{Jonathan Gagn\'e}
\affiliation{\planetarium}
\affiliation{\irex}
\email{gagne@astro.umontreal.ca}

\author[0000-0001-6251-0573]{Jacqueline K. Faherty}
\affiliation{\amnh}
\email{jfaherty@amnh.org}

\author[0000-0001-6733-4118]{Azul Ruiz Diaz}
\affiliation{\planetarium}
\affiliation{\irex}
\email{azul.ruiz.diaz@umontreal.ca}

\author[0000-0002-2195-735X]{Louis-Philippe Coulombe}
\affiliation{\planetarium}
\affiliation{\irex}
\email{louis-philippe.coulombe@umontreal.ca}

\author[0000-0003-2235-761X]{Thomas P. Bickle}
\affil {\openu}
\email{tombicklecrypto@gmail.com}

\author[0000-0002-6294-5937]{Adam C. Schneider}
\affil{United States Naval Observatory, Flagstaff Station, 10391 West Naval Observatory Rd., Flagstaff, AZ 86005, USA}
\email{adam.c.schneider4.civ@us.navy.mil}

\author[0000-0003-4269-260X]{J. Davy Kirkpatrick}
\affil{IPAC, Mail Code 100-22, Caltech, 1200 E. California Blvd., Pasadena, CA 91125, USA}
\email{davy@ipac.caltech.edu}

\author[0000-0002-2387-5489]{Marc J. Kuchner}
\affil{Exoplanets and Stellar Astrophysics Laboratory, NASA Goddard Space Flight Center, 8800 Greenbelt Road, Greenbelt, MD 20771, USA}
\email{marc.j.kuchner@nasa.gov}

\author[0000-0002-1125-7384]{Aaron M. Meisner}
\affil{NSF’s National Optical-Infrared Astronomy Research Laboratory, 950 N. Cherry Ave., Tucson, AZ 85719, USA}
\email{aaron.meisner@noirlab.edu}

\author[0000-0001-7896-5791]{Dan Caselden}
\affiliation{\amnh}
\email{dancaselden@gmail.com}

\author[0000-0002-6523-9536]{Adam J. Burgasser}
\affiliation{Center for Astrophysics and Space Sciences, University of California San Diego, La Jolla, CA 92093, USA}
\email{aburgasser@mail.ucsd.edu}

\author[0000-0003-2478-0120]{Sarah Casewell}
\email{}
\affil{School of Physics and Astronomy, University of Leicester, University Road, Leicester, LE1 7RH, UK}

\author[0000-0003-1202-3683]{Easton J. Honaker}
\email{}
\affil{Department of Physics and Astronomy, University of Delaware, Newark, DE 19716, USA}

\author[0000-0001-8662-1622]{Frank Kiwy}
\email{}
\affil{Backyard Worlds: Planet 9, USA}

\author[0000-0001-7519-1700]{Federico Marocco}
\email{}
\affil{IPAC, Mail Code 100-22, Caltech, 1200 E. California Blvd., Pasadena, CA 91125, USA}

\author[0000-0001-8170-7072]{Daniella C. Bardalez Gagliuffi}
\email{daniellabardalez@gmail.com}
\affiliation{Department of Physics \& Astronomy, Amherst College, Amherst, MA, USA}

\author[0000-0003-4714-3829]{Nikolaj Stevnbak Andersen}
\email{}
\affil{Backyard Worlds: Planet 9, USA}

\author{Lizzeth Ruiz Arroyo}
\email{}
\affil{Backyard Worlds: Planet 9, USA}

\author[0000-0001-8731-9281]{Bruce Baller}
\email{}
\affil{Backyard Worlds: Planet 9, USA}

\author{Paul Beaulieu}
\email{}
\affil{Backyard Worlds: Planet 9, USA}

\author{John Bell}
\email{}
\affil{Backyard Worlds: Planet 9, USA}

\author[0009-0000-5790-7488]{Martin Bilsing}
\email{}
\affil{Backyard Worlds: Planet 9, USA}

\author{Troy K.~Bohling}
\email{}
\affil{Backyard Worlds: Planet 9, USA}

\author[0000-0002-7630-1243]{Guillaume Colin}
\email{}
\affil{Backyard Worlds: Planet 9, USA}

\author[0000-0002-8295-542X]{Giovanni Colombo}
\email{}
\affil{Backyard Worlds: Planet 9, USA}

\author[0009-0004-6814-5449]{Sam Deen}
\email{}
\affil{Backyard Worlds: Planet 9, USA}

\author{Alexandru Dereveanco}
\email{}
\affil{Backyard Worlds: Planet 9, USA}

\author{Kevin Dixon}
\email{}
\affil{Backyard Worlds: Planet 9, USA}

\author[0000-0002-4143-2550]{Hugo A. Durantini Luca}
\email{}
\affil{Backyard Worlds: Planet 9, USA}

\author[0009-0009-0264-1630]{Deiby Flores}
\email{}
\affil{Backyard Worlds: Planet 9, USA}

\author{Christoph Franck}
\email{}
\affil{Backyard Worlds: Planet 9, USA}

\author{Christopher Fulvi}
\email{}
\affil{Backyard Worlds: Planet 9, USA}

\author{Michael Gallmann}
\email{}
\affil{Backyard Worlds: Planet 9, USA}

\author[0000-0002-1044-1112]{Jean Marc Gantier}
\email{}
\affil{Backyard Worlds: Planet 9, USA}

\author{Konstantin Glebov}
\email{}
\affil{Backyard Worlds: Planet 9, USA}

\author[0000-0002-8960-4964]{L\'eopold Gramaize}
\email{}
\affil{Backyard Worlds: Planet 9, USA}

\author[0000-0002-7389-2092]{Leslie K. Hamlet}
\email{}
\affil{Backyard Worlds: Planet 9, USA}

\author[0000-0002-4733-4927]{Ken Hinckley}
\email{}
\affil{Backyard Worlds: Planet 9, USA}

\author{Kevin Jablonski}
\email{}
\affil{Backyard Worlds: Planet 9, USA}

\author[0000-0002-4175-295X]{Peter A. {Ja{\l}owiczor}}
\email{}
\affil{Backyard Worlds: Planet 9, USA}

\author[0000-0003-4905-1370]{Martin Kabatnik}
\email{}
\affil{Backyard Worlds: Planet 9, USA}

\author{Peter Kasprowitz}
\email{}
\affil{Backyard Worlds: Planet 9, USA}

\author[0009-0004-9268-9796]{K Ly}
\email{}
\affil{Backyard Worlds: Planet 9, USA}

\author{David W. Martin}
\email{}
\affil{Backyard Worlds: Planet 9, USA}

\author{Naoufel Marzak}
\email{}
\affil{Backyard Worlds: Planet 9, USA}

\author{Alexander McColgan}
\email{}
\affil{Backyard Worlds: Planet 9, USA}

\author{Neil J.~McEwan}
\email{}
\affil{Backyard Worlds: Planet 9, USA}

\author[0009-0000-8800-3174]{Marianne N. Michaels}
\email{}
\affil{Backyard Worlds: Planet 9, USA}

\author{William Pendrill}
\email{}
\affil{Backyard Worlds: Planet 9, USA}

\author{St{\'e}phane Perlin}
\email{}
\affil{Backyard Worlds: Planet 9, USA}

\author[0000-0001-9692-7908]{Ben Pumphrey}
\email{}
\affil{Backyard Worlds: Planet 9, USA}

\author{James Rabe}
\email{}
\affil{Backyard Worlds: Planet 9, USA}

\author{Henry Raway}
\email{}
\affil{Backyard Worlds: Planet 9, USA}

\author{Walter Ruben Robledo}
\email{}
\affil{Backyard Worlds: Planet 9, USA}

\author{David Roser}
\email{}
\affil{Backyard Worlds: Planet 9, USA}

\author[0009-0005-4611-4008]{Animesh Roy}
\email{}
\affil{Rajshahi University of Engineering \& Technology, Kazla, Rajshahi-6204, Bangladesh}
\affil{Backyard Worlds: Planet 9, USA}

\author[0000-0003-4864-5484]{Arttu Sainio}
\email{}
\affil{Backyard Worlds: Planet 9, USA}

\author[0009-0000-6624-8031]{Vincent Schindler}
\email{}
\affil{Backyard Worlds: Planet 9, USA}

\author{Manfred Schonau}
\email{}
\affil{Backyard Worlds: Planet 9, USA}

\author[0000-0002-7587-7195]{J{\"o}rg Sch{\"u}mann}
\email{}
\affil{Backyard Worlds: Planet 9, USA}

\author{Karl Selg-Mann}
\email{}
\affil{Backyard Worlds: Planet 9, USA}

\author{Andrea Serio}
\email{}
\affil{Backyard Worlds: Planet 9, USA}

\author{Patrick Smith}
\email{}
\affil{Backyard Worlds: Planet 9, USA}

\author{Andres Stenner}
\email{}
\affil{Backyard Worlds: Planet 9, USA}

\author{Christopher Tanner}
\email{}
\affil{Backyard Worlds: Planet 9, USA}

\author[0000-0001-5284-9231]{Melina Th{\'e}venot}
\email{}
\affil{Backyard Worlds: Planet 9, USA}

\author{Vinod Thakur}
\email{}
\affil{Backyard Worlds: Planet 9, USA}

\author[0000-0002-3878-7166]{Mayahuel Torres Guerrero}
\email{}
\affil{Backyard Worlds: Planet 9, USA}

\author{Maurizio Ventura}
\email{}
\affil{Backyard Worlds: Planet 9, USA}

\author{Nikita V. Voloshin}
\email{}
\affil{Backyard Worlds: Planet 9, USA}

\author{Jim Walla}
\email{}
\affil{Backyard Worlds: Planet 9, USA}

\author{Zbigniew W\c edracki}
\email{}
\affil{Backyard Worlds: Planet 9, USA}

\author{Bailey Weyandt}
\email{}
\affil{Backyard Worlds: Planet 9, USA}

\author{Breck Wilhite}
\email{}
\affil{Backyard Worlds: Planet 9, USA}

\author{Spartacus Zitouni}
\email{}
\affil{Backyard Worlds: Planet 9, USA}

\begin{abstract}

We present a Python spectrophotometry extraction tool tailored for fast-moving point sources detected in the SPHEREx mission, and use it to construct a set of 0.75--5.0\,$\mu$m low-resolution ($\lambda/\Delta\lambda \approx 50$) spectrophotometry data products based on the SPHEREx Quick Release 2 (QR2) for a set of \nhqspectra\ L0--Y1 ultracool dwarfs: \nhqknown\ known \added{ultracool} dwarfs, \nphotobd\ known photometric \added{ultracool} dwarf candidates, and \nbrandnewcand\ newly identified ultracool dwarfs. \added{This work more than doubles the number of ultracool dwarfs with spectroscopy, from \nspecucdbefore\ to \nspecucdafter}. We provide SPHEREx templates for each spectral subtype and a set of tools to assign automated spectral types. The QR2 data release generates spectrophotometry with an average signal-to-noise per spectral channel above $\approx$ 10 for most objects with WISE W2 magnitudes of 14.0\,mag and brighter. The compiled data set is made available publicly at \url{https://mocadb.ca}, where new spectral compilations from future data releases will also be made available as they are published. These new data provide a significant increase in the number of substellar objects for which the 2.4--5.0\,$\mu$m window is now accessible, making it possible to probe important molecular chemistry of key CNOS-bearing species for the coolest \added{brown dwarfs}. We flag \nnewyoung\ ultracool dwarfs as candidate young brown dwarfs, \nnewsd\ as candidate subdwarfs, and \nnewpec\ as possibly otherwise peculiar for future investigation. The SPIFF library presented here opens the doors to efficient confirmation of candidate substellar objects and follow-up studies of population-level atmospheric properties of cold brown dwarfs.
\end{abstract}


\section{INTRODUCTION}\label{sec:intro}

The recent advent of the Spectro-Photometer for the History of the Universe, Epoch of Reionization and Ices Explorer (SPHEREx) mission \citep{2026ApJ...999..139B,2020SPIE11443E..0IC,2014arXiv1412.4872D,2018SPIE10698E..1UK} provides unique benefits to the study of ultracool stars and brown dwarfs, even though this was not part of its core mission. SPHEREx launched into low-Earth orbit on 2025, March 11, and began an all sky scan obtaining images with a 3.5\,$\deg$ angular size (consisting of 2048$\times$2048 pixels with angular sizes $\approx 6\farcs15$). The imaging mission began in May of 2025 using a set of 102 narrow spectral channels spanning $\approx 0.75-5.0$\,$\mu$m, with a nominal mission duration of 25 months.

One of the core advantages of SPHEREx for the study of brown dwarfs is the possibility of reconstructing spectrophotometry with properties akin to a low-resolution $\lambda/\Delta\lambda \approx$ 35--125 spectrum \citep{2024SPIE13092E..3NH,2026arXiv260209139H}, with a spectral coverage that includes the 2.4--5.0\,$\mu$m segment, which is difficult to observe from the ground. This wavelength range is particularly important to measure the bolometric luminosities of the coolest brown dwarfs \citep{2015ApJ...810..158F}, and to constrain the atmospheric chemistry of cold, T and Y dwarfs \citep{2006ApJ...648..614C,2006ApJ...648.1181V}, because it contains key molecular bands of CO$_2$ \citep{2010ApJ...722..682Y,2012ApJ...760..151S, 2024Natur.628..511F} and potentially PH$_3$ and SiH$_{4}$ \citep{2006ApJ...648.1181V, 2025Sci...390..697B, 2025Natur.645...62F} that are hard or impossible to constrain at $< $2.4\,$\mu$m alone. The full SPHEREx data also probes molecules such as CO, CH$_4$ and H$_2$O over a wider range of atmospheric pressures. Furthermore, the faint Y dwarfs emit a significant portion of their energy past 2.4\,$\mu$m \citep{2011ApJ...743...50C,2010ApJ...710.1627L,2013ApJ...763..130L}.

However, the data sets produced by the SPHEREx mission are not immediately amenable to this goal of large-scale brown dwarf spectroscopic characterization, because they consist of individual image files that are expensive to download, store, and analyze in order to reconstruct the reduced `spectra' in question. To address this challenge, we present the publicly available `SPHEREx Photometry and Image Fitting Framework' (SPIFF) Python library to reconstruct the spectrophotometry of fast-moving point sources (Section~\ref{sec:spiff}), and use it to build a spectral library of known \added{ultracool} dwarfs (Section~\ref{sec:sample}). We construct \added{hybrid SPHEREx} spectral templates across the 1.0--5.0\,$\mu$m range along with an automated spectral typing scheme, and use it to confirm the substellar nature of \nhqcand\ candidate \added{ultracool} dwarfs (Section~\ref{sec:discussion}). We conclude in Section~\ref{sec:conclusion} with possible areas of research that are unlocked by this new spectral library.

\section{The SPIFF tool}\label{sec:spiff}

\begin{figure*}
	\centering
	\subfigure[PSF reconstruction]{\includegraphics[width=0.95\textwidth]{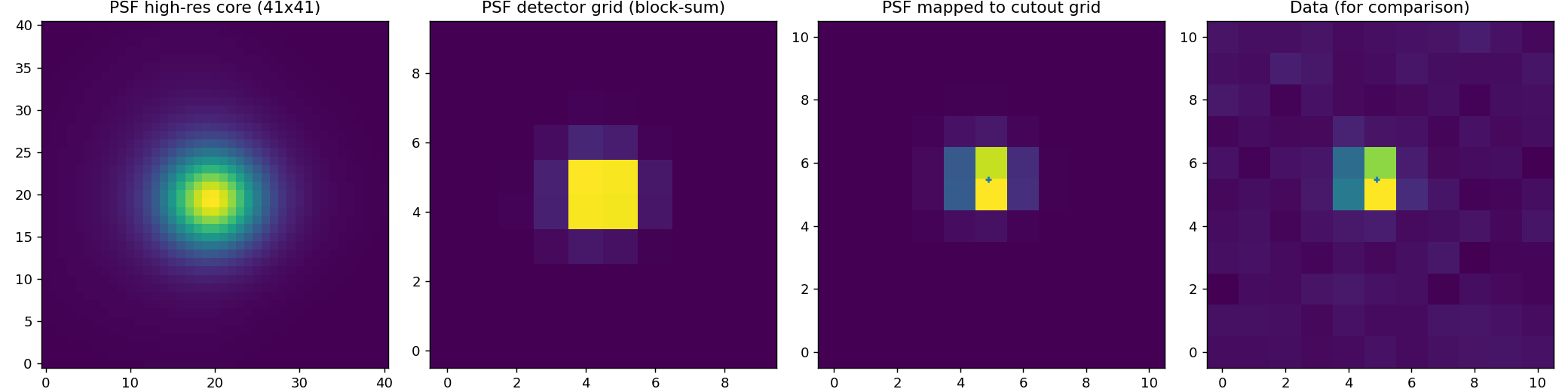}\label{fig:spiff_psf}}
    \subfigure[\texttt{scipy.optimize} PSF fitting]{\includegraphics[width=0.95\textwidth]{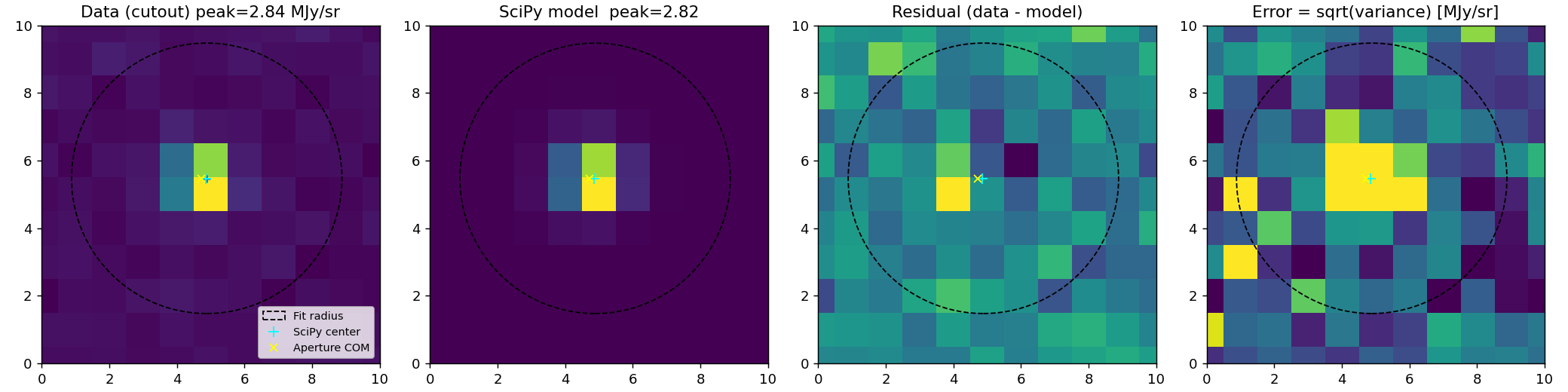}\label{fig:spiff_scipy}}
    \subfigure[Median of \texttt{ultranest} samples]{\includegraphics[width=0.95\textwidth]{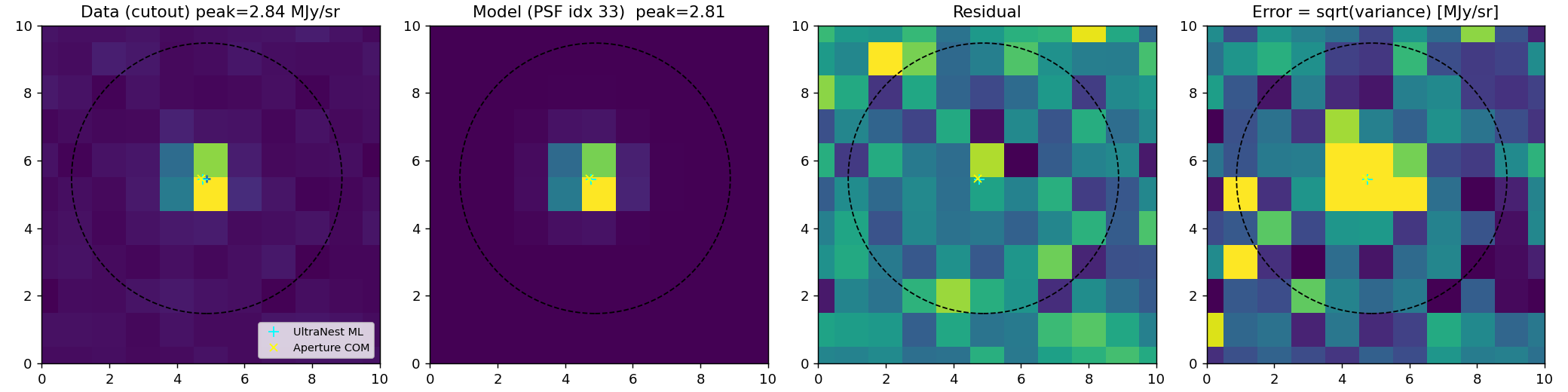}\label{fig:spiff_unest}}
	\caption{Diagnostic figures generated by the SPIFF pipeline. Upper row: PSF reconstruction from the provided over-sampled PSF (left column), placed on the detector grid using a box sum (second column), and mapped onto the observed image (third column). A comparison with an observed PSF for SIMP~J013656.5+093347.3 is shown on the right column. Middle row: PSF fitting using \texttt{scipy.optimize}. Bottom row: Final PSF fitting parameters obtained from the median samples of \texttt{ultranest}. See Section~\ref{sec:spiff} for more details.}
	\label{fig:spiff_fitting}
\end{figure*}

The `SPHEREx Photometry and Image Fitting Framework' (SPIFF) Python tool is analogous to the spectrophotometry tool \citep{2025arXiv251115823A} provided by the NASA IPAC InfraRed Science Archive (IRSA) \footnote{Available at \url{https://irsa.ipac.caltech.edu/applications/spherex/tool-spectrophotometry}.}, with a few different characteristics. The IRSA tool takes as input several coordinates on which to perform forced photometry using the SPHEREx instrumental point spread function (PSF) appropriate for the detector position for a given visit, as provided in the SPHEREx calibration files. Background subtraction is performed locally by subtracting the median of the flux in a 15-pixel-wide box, after masking bad pixels and known stars. The spectrophotometry of the target sources can then be extracted, accounting for crowding effects. 

The IRSA spectrophotometry tool requires manual input of the target coordinates, and does not allow for large batch queries, nor does it account for the proper motion of objects that move significantly between SPHEREx epochs. The latter constraint can be problematic for fast-proper motion objects, especially given that the tool performs forced photometry without allowing the exact PSF position to move during the fitting procedure. This makes it particularly important to provide precise coordinates for proper results.

In order to extract the spectrophotometry of a large number of fast-moving substellar objects efficiently, and to do so routinely as the SPHEREx data get released in increments, we built a custom \texttt{Python} SPHEREx spectrophotometry tool, the `SPHEREx Photometry and Image Fitting Framework' (SPIFF)\footnote{Available at \url{https://github.com/jgagneastro/SPIFF}.}. This tool works similarly to the IRSA spectrophotometry tool, with the following notable differences:
\begin{itemize}
    \item The proper motion of the target can be provided such that the initial detector position of the PSF is adjusted using the epoch of the individual SPHEREx image\footnote{\added{We do not attempt to model parallax motion because it is negligible compared to a SPHEREx pixel even for the nearest brown dwarf.}}; 
    \item only a single PSF can be provided;
    \item the detector position of the PSF is allowed to vary during the PSF fitting step;
    \item the PSF fitting step is performed with \texttt{scipy.optimize}, and the Bayesian likelihood corresponding to measurement uncertainties is explored with the \texttt{ultranest} nested sampling package.
\end{itemize}

Examples of PSF fitting steps are provided in Figure~\ref{fig:spiff_fitting}. Similar to the IRSA spectrophotometry tool, SPIFF uses the instrumental SPHEREx PSF provided in the calibration files, which are oversampled with respect to detector pixels, and performs a lateral shift using linear interpolation before downgrading the spatial resolution of the PSF onto the spatial resolution of the observed image using a box sum (after shifting the PSF). An example of this step is provided in Figure~\ref{fig:spiff_psf}.

The SPIFF pipeline avoids downloading the full 2048$\times$2048-pixels FITS image or each data point to improve efficiency. Instead, it first uses the IPAC SIA2 service to identify the individual URLs of FITS images in a 5\arcsec\ cone search around the target position projected to \added{epoch} 2025.5. For each individual FITS image, the SPIFF pipeline then requests a small 20$\times$20-pixels cutout of the FITS image stored on Amazon S3, around the coordinates of the target projected to the epoch of the specific FITS image observation. This substantially reduces the network transfers required for analysis, but downloading the headers of the FITS images can still be the bottleneck in terms of total computing time required to construct a SPIFF data product, depending on internet bandwidth. In order to run SPIFF efficiently on large samples, we utilized the Digital Research Alliance resources\footnote{\url{alliancecan.ca}}, which allow for both fast computing and high internet bandwidth.

The data products of the SPIFF pipeline consist of quality flags, extracted central wavelength, approximate wavelength bin width (obtained from the SPHEREx calibration files at the best-fitting PSF detector position), as well as the flux and measurement error provided in three flavors:
\begin{itemize}
    \item using aperture photometry with a set of 18 aperture radii (1.0-5.0 pixels, $6\farcs5$--$30\farcs8$), preserving the measurement that maximizes S/N,
    \item based on the scipy optimization of the PSF fitting,
    \item and based on the median of the ultranest samples (for the measurement) and the median absolute deviation (for the measurement error) when exploring the PSF fitting likelihood.
\end{itemize}

The full columns of the SPIFF outputs are described in Table~\ref{tab:spiff}. In practice, the most reliable flux measurements are usually obtained from the median ultranest samples, because it tends to explore the likelihood more thoroughly than \texttt{scipy.optimize}, and it is much less sensitive to background contaminants compared with aperture photometry, especially given the large $6\farcs15$ pixel size of the SPHEREx detectors.

In this work, we present a set of \nhqspectra\ high-quality spectrophotometry products compiled with SPIFF, and stored in the Montreal Open Clusters and Associations (MOCA) database\footnote{Available at \url{https://mocadb.ca}. A persistent and versioned copy of the SPIFF spectrophotometric library is also available on Zenodo at \href{https://doi.org/10.5281/zenodo.19051216}{doi:10.5281/zenodo.19051216}.} \citep{2024PASP..136f3001G,2026arXiv260215695G}. They are stored as raw SPIFF outputs in the table \texttt{data\_spherex\_spectra\_spiff}. However, SPHEREx revisits each sky location multiple times, and every time the target falls on a slightly different region of the detector, resulting in small wavelength shifts. Because every such measurement is actually a photometric measurement with a relatively wide 0.02--0.08\,$\mu$m bandpass \citep{2026arXiv260209139H}, the true spectral resolving power is not as high as these repeated measurements with small wavelength shifts may suggest. Instead, this results in an oversampling of the spectral resolution element. We also provide a simpler data product in the MOCAdb \texttt{data\_spectra} table, in which we compile the error-weighted mean of all available flux measurements in each of the 102 SPHEREx spectral channels. The current version of SPIFF relies on preliminary response functions of each SPHEREx spectral channel, detailed in SPHEREx mission's Quick Release (QR) explanatory supplements\footnote{Available at \url{https://irsa.ipac.caltech.edu/data/SPHEREx/docs/SPHEREx_Expsupp_QR.pdf}.}, listed in Table~\ref{tab:bandpass}, although refined response functions will eventually be released \citep{2026arXiv260209139H} and implemented in future versions of SPIFF. It is worth noting that each individual pixel of the SPHEREx detectors have their own specific response functions, which can vary across the detector even with a fixed spectral channel, meaning that the reconstructed spectrophotometry stored in \texttt{data\_spherex\_spectra\_spiff} are spectrally over-sampled in a complex way, and the binned spectra provided in the \texttt{data\_spectra} table are an approximate reconstruction. Example MySQL queries to recover the raw measurements or the compiled spectra are provided in the Appendix.

\section{SAMPLE}\label{sec:sample}

\begin{figure*}
	\centering
	\subfigure[Spectral type distribution of our sample]{\includegraphics[width=0.48\textwidth]{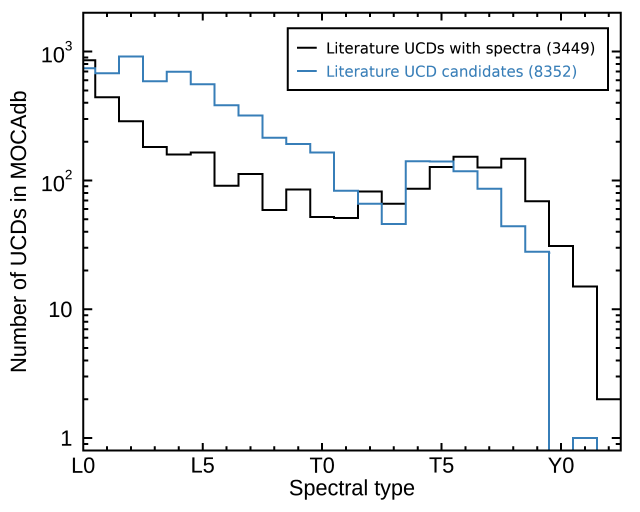}}\label{fig:nspectra_spt_sample}
    \subfigure[WISE W2 magnitudes of our sample]{\includegraphics[width=0.48\textwidth]{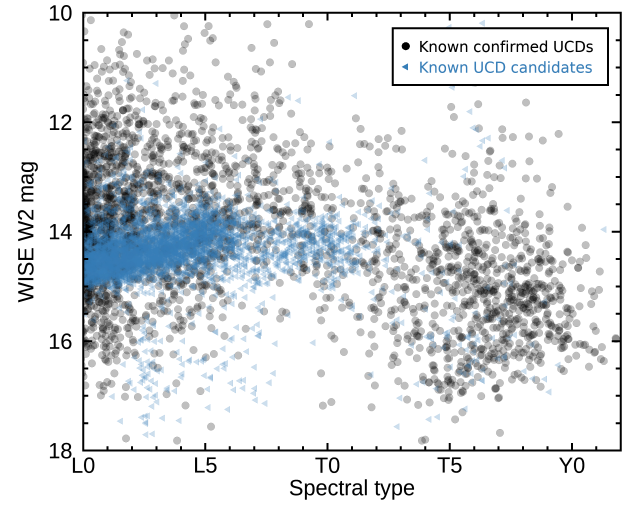}}\label{fig:spt_w2_sample}
	\caption{Left panel: Histogram of MOCAdb spectral types for known confirmed and candidate ultracool dwarfs. Right panel: WISE W2 magnitudes as a function of spectral types for known confirmed and candidate ultracool dwarfs. \added{Known UCD candidates are concentrated at early L spectral types and magnitudes brighter than W2$\approx$15 because they were mostly uncovered using the individual-epoch detections in the WISE mission (which reaches a 5$\sigma$ depth of 15.5 in W2 band \citealp{2010AJ....140.1868W}), which samples a much larger volume in space for warmer and intrinsically brighter early-L dwarfs.} See Section~\ref{sec:sample} for more details.}
	\label{fig:sample}
\end{figure*}

We applied the SPIFF pipeline on the set of all known ultracool dwarfs (UCDs), consisting mostly of brown dwarfs with some older very low-mass stars (spectral types L0 to $\approx$\,L2, \citealp{2000AJ....120.1085G,2014AJ....147...94D}), maintained as part of the MOCA database \citep{2024PASP..136f3001G,2026arXiv260215695G}, which benefited from the Ultracoolsheet \citep{Best20US}\footnote{A persistent and versioned copy of the Ultracoolsheet is available on Zenodo at \href{https://doi.org/10.5281/zenodo.4169084}{doi:10.5281/zenodo.4169084}}, and the SIMPLE archive \cite{zenodosimplearchive}\footnote{\added{A persistent and versioned copy of the SIMPLE archive is available on Zenodo at \href{https://doi.org/10.5281/zenodo.13937301}{doi:10.5281/zenodo.13937301}, and the current version is available at \url{https://simple-bd-archive.org}.}}. \added{In this work, we define ultracool dwarfs as all objects with spectral types L0 and later.} We considered all spectroscopically confirmed \added{ultracool} dwarfs and selected the best-available reference astrometry and proper motion from the literature. This set of literature \added{ultracool} dwarfs is available in the individual tables of \citet{2026arXiv260215695G}\footnote{A persistent and versioned copy of the MOCAdb is available on Zenodo at \href{https://doi.org/10.5281/zenodo.18166117}{doi:10.5281/zenodo.18166117}}. The spectral type and W2 magnitude distribution of this sample are shown in Figure~\ref{fig:sample}.

\added{In addition to the set of known ultracool dwarfs, we compiled photometrically-selected ultracool dwarf candidates from the literature and custom searches in wide-area red and infrared surveys, to attempt a confirmation of their ultracool nature using SPHEREx spectrophotometry.} The \added{ultracool} dwarf candidates were compiled from 4 different samples:

\begin{itemize}
    \item All SIMBAD \citep{2000AAS..143....9W} objects flagged as brown dwarf candidates,
    \item the \added{ultracool} dwarf candidates identified through the Backyard Worlds: Planet~9 citizen science project (\citealp{2017ApJ...841L..19K,2026arXiv260401323S}),
    \item the BASS \citep{2015ApJS..219...33G} and BASS-Ultracool \citep{2015ESS.....310419G} searches for young ultracool dwarfs,
    \item \added{a search based on colors and proper motions (when available) in CatWISE \citep{2021ApJS..253....8M}, UKIDSS~LAS~DR11$+$ \citep{2007MNRAS.379.1599L}, UHS~DR3 \citep{2025AJ....170...86S},Gaia~DR3 \citep{2023AA...674A...1G} and Pan-STARRS1~DR2 \citep{2016arXiv161205560C} with colors consistent with L0 or later in each detected spectral band\footnote{See \citealp{2011ApJS..197...19K,2018ApJS..234....1B,2010MNRAS.406.1885B} for average \added{ultracool} dwarf colors in each respective catalogs.}}
\end{itemize}

\added{The UHS~DR3 selection criteria were designed in a way to recover fast-moving point sources in a single and efficient ADQL query:}
\begin{itemize}
  \item $\sigma_{\mu_{\alpha}}$ and $\sigma_{\mu_{\alpha}} < 50$\,\masyr,
  \item $|\mu_{\alpha}|$ or $|\mu_{\delta}| \ge 50$\,\masyr,
  \item at least one proper motion component is significant at the $5\sigma$ level,
  \item $P_{\mathrm{saturated}} < 0.1$,
  \item $P_{\mathrm{star}} > 0.9$,
  \item No Gaia~EDR3 \citep{2021AA...649A...1G} counterpart exists within $8''$, based on the \texttt{uhsSourceXGEDR3gaia\_source} table at \url{http://wsa.roe.ac.uk}.
\end{itemize}

\added{We used a Gaia~DR3 rejection criterion in UHS~DR3 because it was readily available in its database schema and is efficient at removing earlier-type contaminants from the sample. The Gaia UCDs are more easily recovered with a specific search based on the Gaia catalog directly.}

\added{The criteria we used for CatWISE were the following:}
\begin{itemize}
  \item $\mu_{\rm tot} \geq $20\,\masyr,
  \item $\mu_{\rm tot}$ is detected at 5$\sigma$,
  \item $-0.5 < \mathrm{w1sky} < 0.5$,
  \item $-0.5 < \mathrm{w2sky} < 1$,
  \item $W1 - W2 \geq 0.5$,
  \item $\sigma_{W2}$ must be detected with $\sigma_{W2} \leq 0.15$.
\end{itemize}

\added{The criteria for UKIDSS~LAS~DR11$+$ were:}
\begin{itemize}
  \item $|\mu_{\alpha}| + |\mu_{\delta}| \ge 100$\,\masyr,
  \item $Y - J \ge 1.7$ if $Y$ is detected with $\sigma_{y} < 0.3$,
  \item $J$ msut be detected with $\sigma_{J} < 0.3$,
  \item $P_{\mathrm{star}} \ge 0.8$.
\end{itemize}

\added{The criteria for Gaia~DR3 were:}
\begin{itemize}
    \item The parallax is $\varpi \ge 10$\,mas,
    \item $\varpi / \sigma_{\varpi} \ge 10$,
    \item \texttt{visibility\_periods\_used} $\ge 10$,
    \item \texttt{ruwe} $< 1.4$,
    \item \texttt{duplicated\_source} = \texttt{false},
    \item \texttt{phot\_rp\_mean\_flux\_over\_error} $> 10$,
    \item \texttt{phot\_g\_mean\_flux\_over\_error} $> 10$,
    \item $G - G_{\mathrm{RP}} \ge 1.55$,
    \item Approximate absolute $G$-band magnitude $G + 5\log_{10}(\varpi) - 10 \ge 14.0$.
\end{itemize}

\added{The criteria for Pan-STARRS1~DR2 were:}
\begin{itemize}
  \item $y$ and $z$ must be detected with $\sigma_{y} < 0.1$,
  \item The source must lie outside the Galactic plane: $|b| > 15^\circ$
  \item $g-z \geq 5$ if $\sigma_{g} < 0.5$,
  \item $r-z \geq 3.5$ if $\sigma_{r} < 0.5$,
  \item $i-z \geq 1.5$ if $\sigma_{i} < 0.5$,
  \item $z-y \geq 1$.
\end{itemize}

\section{DISCUSSION}\label{sec:discussion}

The application of the SPIFF pipeline to the samples described in Section~\ref{sec:sample} resulted in \nhqspectra\ spectrophotometry products, which include every available epochs from the SPHEREx mission as of \completiontimestamp. We refer to this data set as the SPIFF spectral library\footnote{\added{A persistent and versioned copy of the SPIFF spectrophotometric library is available on Zenodo at \href{https://doi.org/10.5281/zenodo.19051216}{doi:10.5281/zenodo.19051216}}}.

\begin{figure*}
	\centering
	\subfigure[Differences between the SPIFF and IRSA data reduction pipelines]{\includegraphics[width=0.5\textwidth]{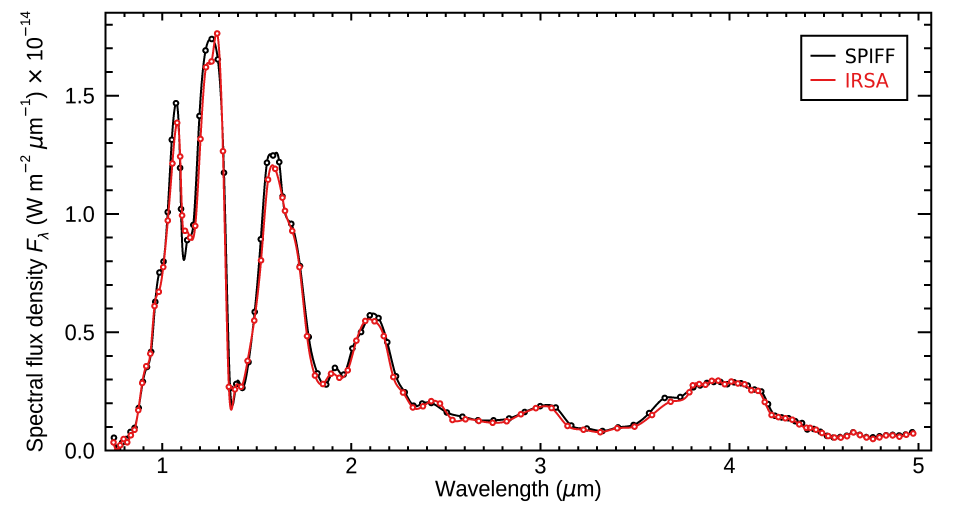}\label{fig:spiff_irsa}}
    \subfigure[Differences between SPIFF and JWST-based synthetic SPHEREx spectrophotometry]{\includegraphics[width=0.5\textwidth]{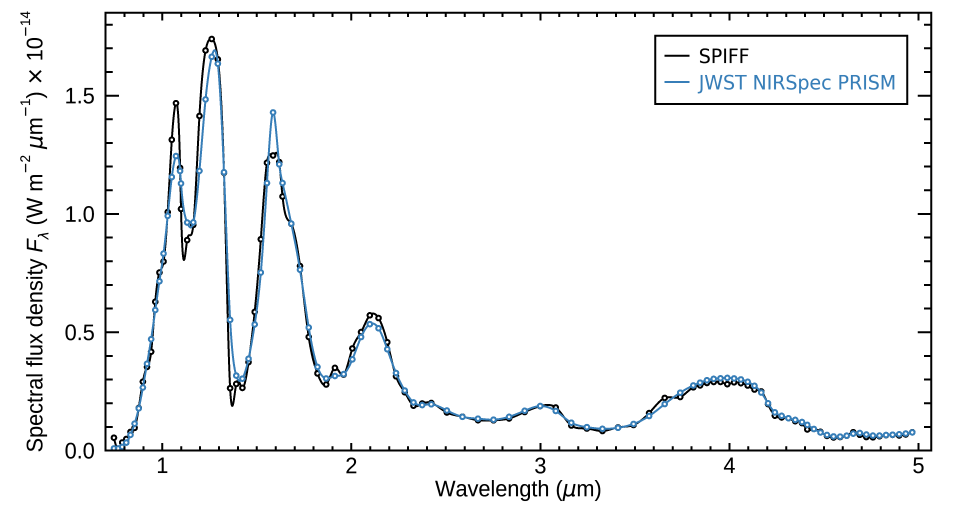}\label{fig:spiff_jwst}}
    \subfigure[Differences between IRSA and JWST-based synthetic SPHEREx spectrophotometry]{\includegraphics[width=0.5\textwidth]{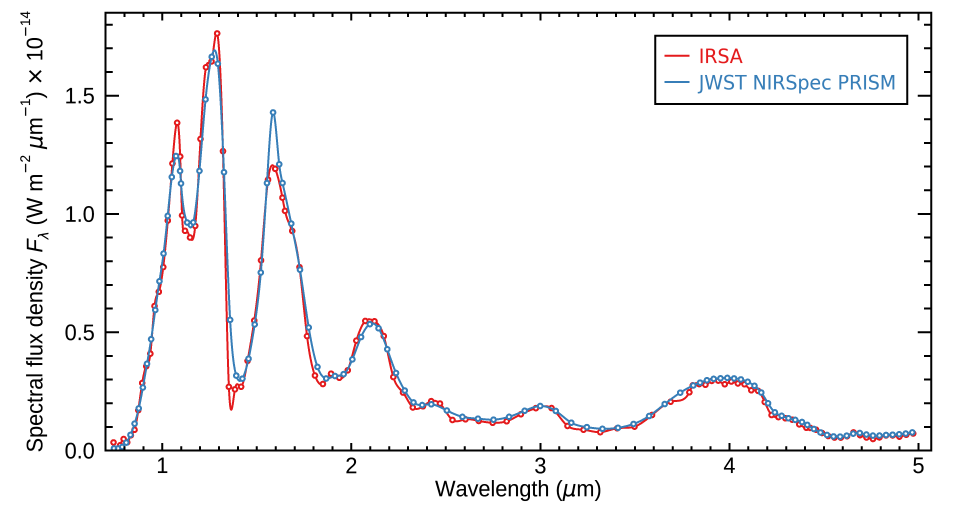}\label{fig:irsa_jwst}}
	\caption{Differences in extracted absolute flux between the SPIFF and IRSA spectrophotometry tools for the T2.5 substellar object SIMP~J013656.5+093347.3. Synthetic spectrophotometry was also reconstructed based on the JWST NIRSpec prism observations of GO program 3548, where the median of all observed epochs is shown for each wavelength bin. The SPIFF pipeline recovers absolute flux values slightly more consistent with those of JWST (the median of residuals indicate a 2.2\% smaller flux in the SPIFF reduction, with median absolute deviations of 7.6\%, whereas the residuals of the IRSA spectrum are 4.9\% fainter with a median absolute deviation of 8.6\%). Splines are shown as solid lines to guide the eye only. See Section~\ref{sec:validation} for more details.}
	\label{fig:spiff_vs_irsa_vs_jwst}
\end{figure*}

\subsection{Pipeline Validation}\label{sec:validation}

We used the JWST GO program 3548 and the IRSA spectrophotometry tool to assess the absolute flux calibration of the SPIFF library (see Figure~\ref{fig:spiff_vs_irsa_vs_jwst}) based on the nearby, bright T2.5 dwarf SIMP~J013656.5+093347.3 (SIMP~J0136; \citealp{2006ApJ...651L..57A}), observed with JWST/NIRSpec using the PRISM/CLEAR configuration with the S1600A1 slit and SUB512 subarray in Bright Object Time-Series mode \citep{2025ApJ...993..237A}. We choose this particular data set for comparison because SIMP~J0136 is a well-characterized, bright source representative of our sample for which we obtain high-S/N SPHEREx spectrophotometry, with a well-characterized astrometry (crucial for the IRSA spectrophotometry tool) and variability (1.0--2.5\% in the 1--5\,$\mu$m range; \citealp{2025ApJ...981L..22M}). The GO3548 observations also cover the same wavelength range over a range of epochs. The current JWST NIRSpec documentation\footnote{See \url{https://jwst-docs.stsci.edu/jwst-calibration-status/nirspec-calibration-status/nirspec-fixed-slit-calibration-status}.} notes:

Current fixed slit absolute flux calibration is generally better than 5\% absolute for sources well centered in the aperture, and should be good to $\approx$\,3\% for most disperser-filter combinations (2\% relative). An exception is for observations in S1600A1 [which applies to our data set] at longer than 3.5\,$\mu$m, where, with the current F-flat reference files \citep{2016SPIE.9904E..46R}, the current absolute flux accuracy can be 5\%--6\%. New F-Flats are expected to be delivered in the near future that should provide absolute flux calibration good to 1\%--2\%.'

Given these flux calibration requirements, the photometric variability of SIMP~J0136 should not affect our assessment of flux calibration. The JWST data\footnote{Obtained from MAST at \url{https://archive.stsci.edu/missions-and-data/jwst}, with the provided reduced spectra based on \texttt{SDP\_VER=2025\_4a, PRD\_VER=PRDOPSSOC-072, OSS\_VER=008.007.000.000, GSC\_VER=GSC31, CAL\_VER=1.20.2, CAL\_VCS=RELEASE, CRDS\_VER=13.0.6, CRDS\_CTX=jwst\_1464.pmap}.} were convolved on the SPHEREx bandpasses listed in Table~\ref{tab:bandpass}. This particular data set covers a range of epochs, and we adopted the median of all epochs along with the median absolute deviation as our reference flux and flux error.

The SPHEREx mission design aims for an absolute flux calibration of 2\%\ within each spectral channel \citep{2024AAS...24433206M,2024AAS...24336002M}, although the empirical in-flight performance has not yet been published (\citealp{2024SPIE13092E..3NH}; Ashby et al., in prep.).

We performed our comparison using the IRSA and SPIFF spectrophotometry products grouped in the 102 spectral channels, as described in Section~\ref{sec:spiff}. We find only slight differences in absolute flux calibration: SPIFF recovers a median absolute flux 2.2\%\ fainter than the JWST spectrum, but 3.4\%\ brighter than the IRSA spectrophotometry, with respective median absolute deviations of the residuals of 7.6\%\ and 5.8\%\ when comparing SPIFF and JWST or IRSA and JWST, respectively. The differences between the SPIFF and IRSA results are based on the same data products and SPHEREx calibration, and must therefore be related either to different performances in the PSF-fitting algorithm, or to a slight mismatch between the provided astrometry and the true location of the PSF on the SPHEREx detectors. It is plausible that the thorough likelihood exploration provided by \texttt{ultranest}, combined with the flexible PSF detector position, may allow slightly more flux to be recovered than the IRSA pipeline. The differences between SPIFF and JWST are close to the expected NIRSpec absolute flux calibration performance, and may also be related to the current calibration of the SPHEREx data products that are still being improved (especially background subtraction) and to our use of preliminary SPHEREx bandpasses to convolve the JWST spectrum (we expect these differences to decrease as future SPHEREx data releases occur). Based on these comparisons, we recommend adopting a conservative error floor of $\approx$\,8\%\ on the absolute flux calibration measurements provided by SPIFF, \added{until a more thorough characterization of SPHEREx absolute flux calibration is available}. We do not apply this floor directly in the SPIFF data products, because measurement errors smaller than this floor may still be a valid representation of relative flux errors.

\begin{figure*}
	\centering
	\subfigure[All high-quality SPHEREx spectra presented here]{\includegraphics[width=0.48\textwidth]{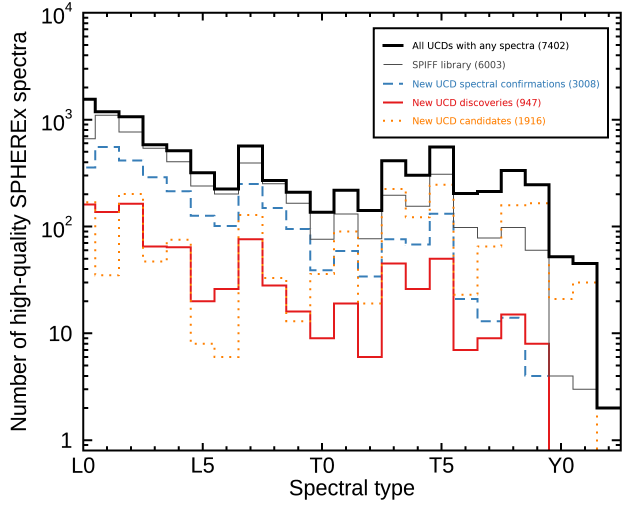}}\label{fig:nspectra_spt}
    \subfigure[Average SPHEREx S/N versus WISE W2 magnitude]{\includegraphics[width=0.48\textwidth]{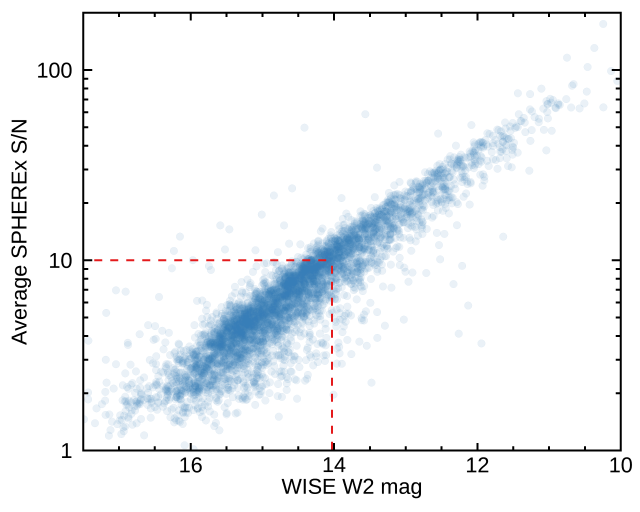}}\label{fig:snr_w2}
	\caption{Left panel: Histogram of high-quality SPHEREx spectrophotometry products presented here. The thick black line represents the current population of ultracool dwarfs with spectral type determinations based on spectroscopy, including the SPIFF library (which is shown independently as a thin gray line)\added{, as well as every other spectrometer}. A significant fraction of known and newly-discovered ultracool dwarfs with spectral types T5 and earlier have spectrophotometry data products available in the SPIFF library. New spectroscopic confirmations of literature ultracool dwarf candidates are shown with a dashed blue line. New discoveries of confirmed or candidate ultracool dwarfs are shown with red and dotted orange lines, respectively. Right panel: Average spectrophotometry S/N as a function of WISE W2 magnitude. Objects with W2 $=14$ or brighter usually result in high-quality SPHEREx spectrophotometry with an average S/N of 10 or above, when free from background contamination.  See Section~\ref{sec:validation} for more details.}
	\label{fig:snr}
\end{figure*}

In Figure~\ref{fig:snr}, we show the average signal-to-noise ratio (S/N) for the SPIFF spectrophotometry as a function of WISE W2 magnitudes across the sample of known \added{ultracool} dwarfs, along with a histogram of the spectral types for the high-quality SPHEREx spectra. We find that data products with an average S/N of 10 per spectral channel correspond to a WISE W2 magnitude of \wmagsnr\,mag, on average. We observed only a very small dependency of the average S/N on the ecliptic latitude of the targets. SPHEREx schedules more frequent visits near the ecliptic poles which increases the number of data points per spectral channel, but the impact on the average S/N of the data binned onto the 102 SPHEREx spectral channels is much smaller than that of the W2 magnitude of the target.

\subsection{Hybrid Spectral Templates}\label{sec:sptt}

We built hybrid spectral templates for each spectral subtype in the range M9--Y1 by combining the available SPHEREx spectra (binned into the 102 SPHEREx spectral channels) in our library for each known \added{ultracool} dwarf of the corresponding spectral type in the literature, while ignoring those with mentions of peculiarities (including low gravity, subdwarfs and spectral binaries). Each spectrum was scaled to its median before constructing a first template estimate based on a per-spectral channel median of each spectrum, and then the scaling that optimizes the $\chi^2$ between each individual spectrum and the template estimate was computed to scale each spectrum individually in an optimal manner. After this was completed, the individual spectra were combined again, using a median. The median absolute deviation was also recorded in each spectral channel to represent the typical spread within a spectral subtype while ignoring large outliers.

\begin{figure*}
	\centering
	\subfigure[L0 SPHEREx template based on 607 objects]{\includegraphics[width=0.48\textwidth]{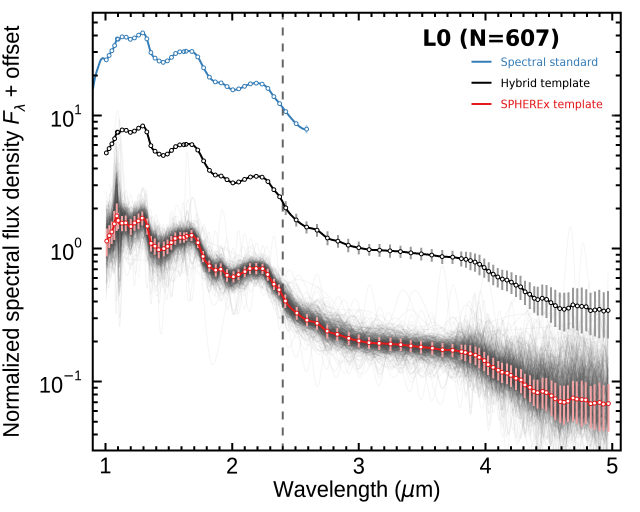}\label{fig:template_l0}}
    \subfigure[L5 SPHEREx template based on 112 objects]{\includegraphics[width=0.48\textwidth]{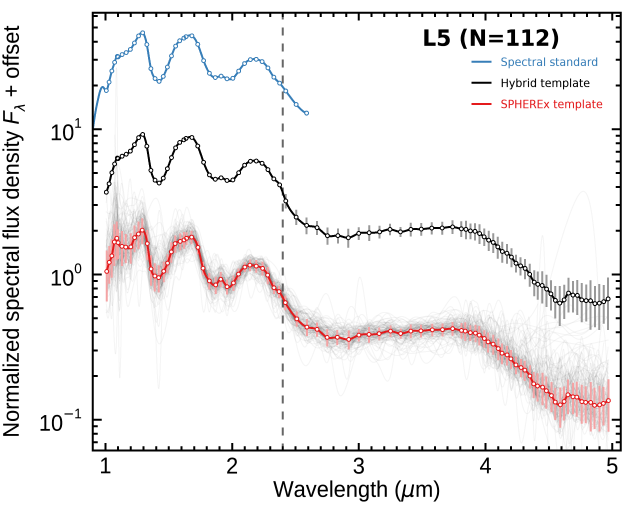}\label{fig:template_l5}}
    \subfigure[T0 SPHEREx template based on 32 objects]{\includegraphics[width=0.48\textwidth]{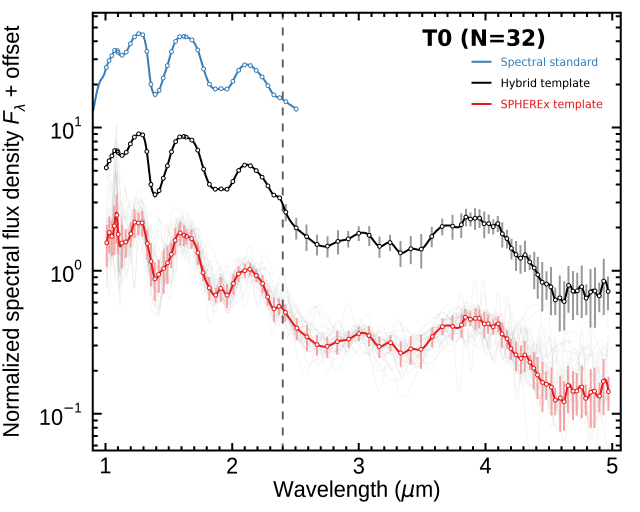}\label{fig:template_t0}}
    \subfigure[T5 SPHEREx template based on 84 objects]{\includegraphics[width=0.48\textwidth]{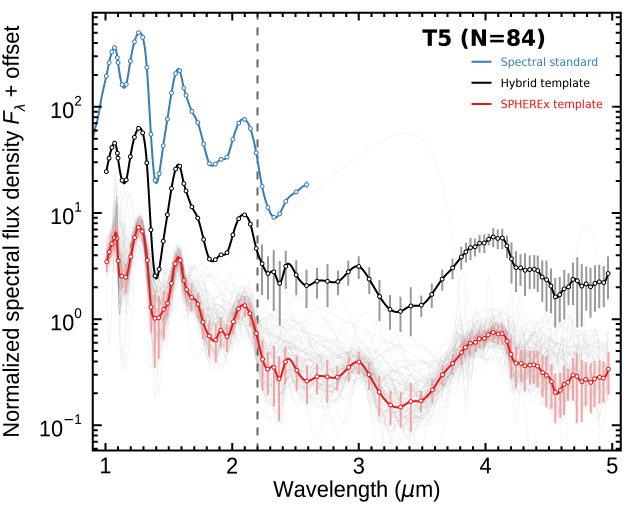}\label{fig:template_t5}}
	\caption{Examples of \added{hybrid} SPHEREx template construction. Synthetic spectrophotometry from a known spectral standard is shown in blue (top), and a raw SPHEREx template constructed from the median of all high-quality SPHEREx data with the appropriate spectral type are shown in red (bottom). The hybrid template, constructed by extending the synthetic standard-based spectrophotometry with the raw SPHEREx template beyond 2.4\,$\mu$m (vertical, dashed line) is shown in black (middle row). For some templates, this limit was moved to 2.1--2.3\,$\mu$m when the quality of the ground-based data was insufficient in this region. The gray error bars do not represent the error on the mean, but the median absolute deviation of all observed SPHEREx spectra to represent visually the typical spread in observed SPHEREx fluxes, which are mostly due to the limited S/N of the spectrophotometry. Splines are shown as solid lines to guide the eye only. The complete figure set (\ntemplates\ images) is available in the online journal. See Section~\ref{sec:sptt} for more details.}
	\label{fig:sptt}
\end{figure*}

A spectral standard from the literature (see Table~\ref{tab:std}) was then convolved onto the set of SPHEREx bandpasses to create a synthetic SPHEREx spectrum at wavelengths shorter than $\approx 2.4$\,$\mu$m. \added{A linear error propagation formula was used to determine the measurement error within each SPHEREx spectrophotometric band.} We then determined the scaling factor that minimized $\chi^2$ with our template, and replaced all channels blue than 2.4\,$\mu$m in our \added{raw} SPHEREx template with the synthetic standard, which results in a higher S/N, especially for the few spectral classes with only a small sample of high-quality SPHEREx spectra. \added{We refer to these templates as hybrid in the remainder of this work, and we also make the raw SPHEREx templates available in the online-only material of this paper.} Figure~\ref{fig:sptt} shows example \added{hybrid} template construction steps\added{; we note that the displayed error bars do not represent the error on the mean, but the median absolute deviation of all observed SPHEREx spectra, dominated by the limited S/N of the individual SPHEREx spectrophotometry products}.

\begin{figure*}[!t]
	\centering
	\includegraphics[width=0.9\textwidth]{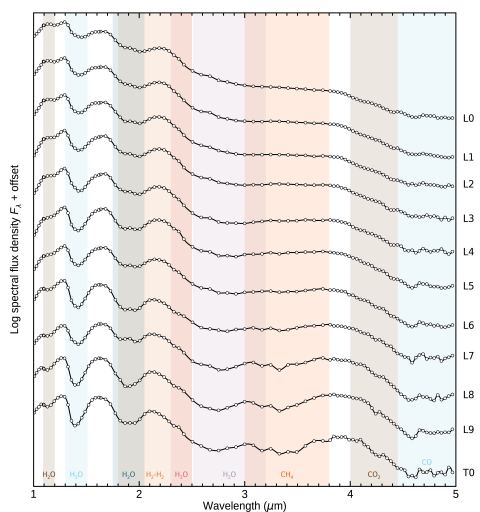}
	\caption{Sequence of field L-type SPHEREx field \added{hybrid} templates constructed in this work, along with relevant chemical species. Splines are shown as solid lines to guide the eye only. See Section~\ref{sec:sptt} for more details.}
	\label{fig:l_templates}
\end{figure*}

\begin{figure*}[!t]
	\centering
    \includegraphics[width=0.9\textwidth]{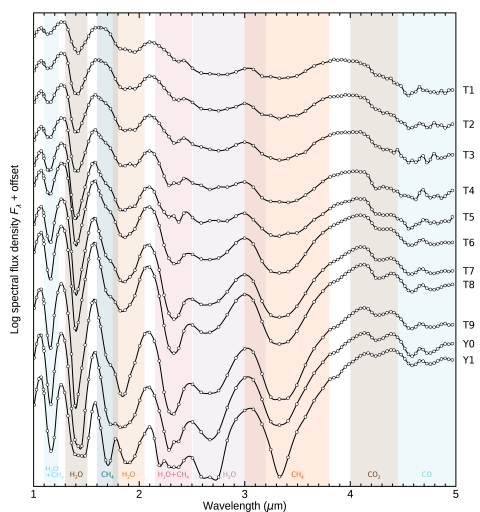}
	\caption{Sequence of field T- and Y-type SPHEREx field \added{hybrid} templates constructed in this work, along with relevant chemical species. Splines are shown as solid lines to guide the eye only. See Section~\ref{sec:sptt} for more details.}
	\label{fig:ty_templates}
\end{figure*}

For spectral types T6--Y1, we used JWST NIRSpec PRISM spectra from programs GO~2302, GTO~1189, and GTO~4539 for the brown dwarfs SDSSp~J162414.37+002915.6 (T6, \citealp{2025ApJ...982...79B}), 2MASS~J03480772-6022270 (T7, \citealp{2006ApJ...637.1067B}), WISEPC~J225540.74-311841.8 (T8, \citealp{2011ApJS..197...19K}), WISE~J210200.15-442919.5 (T9, \citealp{2012ApJ...753..156K}), WISEPC~J205628.90+145953.3 (Y0, \citealp{2011ApJS..197...19K}), and WISEPA~J154151.66-225025.2 (Y1, \citealp{2015ApJ...804...92S}) as they cover the full range of SPHEREx bandpasses and the available SPHEREx spectrophotometry are of a much lower S/N. The final set of field \added{hybrid} templates is shown in Figures~\ref{fig:l_templates} and \ref{fig:ty_templates}.

These steps were repeated \added{to construct hybrid templates} for low-gravity and subdwarf \added{standards} available in the literature (also in Table~\ref{tab:std}) when either the literature data or available SPHEREx data products allowed a full wavelength coverage from 1\,$\mu$m to 5\,$\mu$m.

\subsection{Spectral Typing}\label{sec:spt}

\begin{figure}
	\centering
	\includegraphics[width=\linewidth]{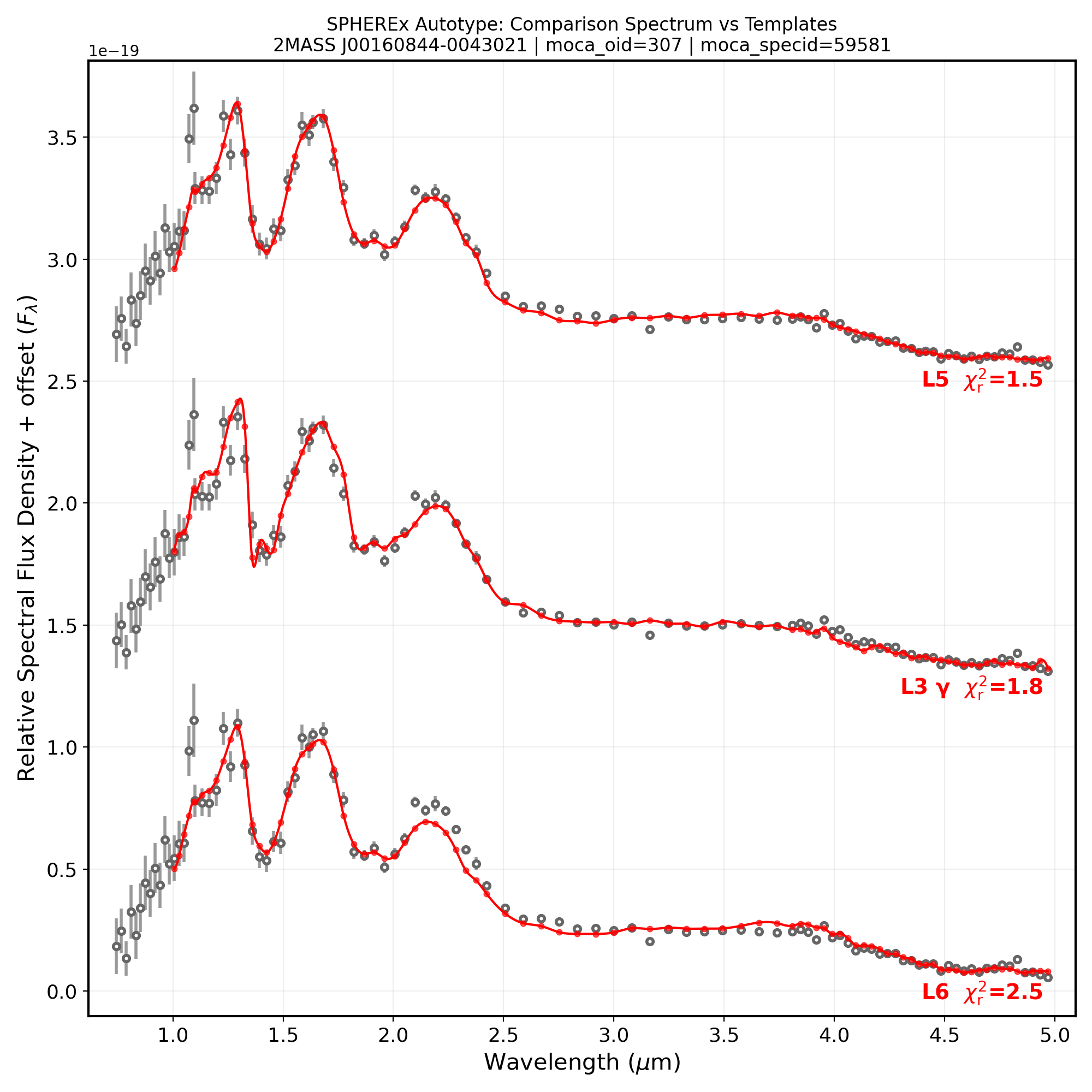}
	\caption{Template-fitting validation figures for the high-quality spectra of literature \added{ultracool} dwarfs. The complete figure set (\nhqknown\ images) is available in the online journal. See Section~\ref{sec:spt} for more details.}
	\label{fig:spt_fitting}
\end{figure}

\begin{figure*}[!t]
	\centering
    \subfigure[A contaminating background star]{\includegraphics[width=0.47\textwidth]{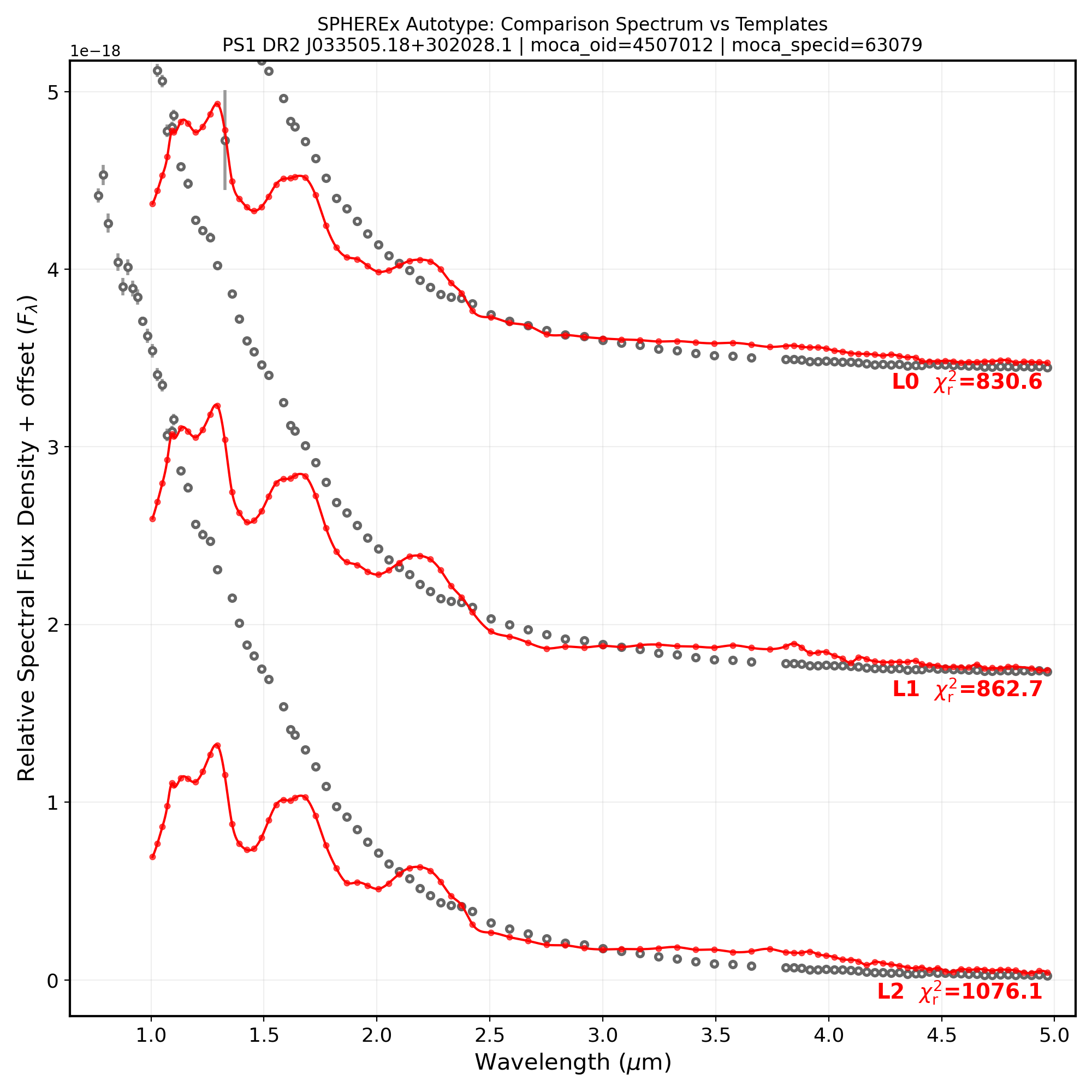}\label{fig:autotypestar}}
    \subfigure[A false-positive star reddened by interstellar dust]{\includegraphics[width=0.47\textwidth]{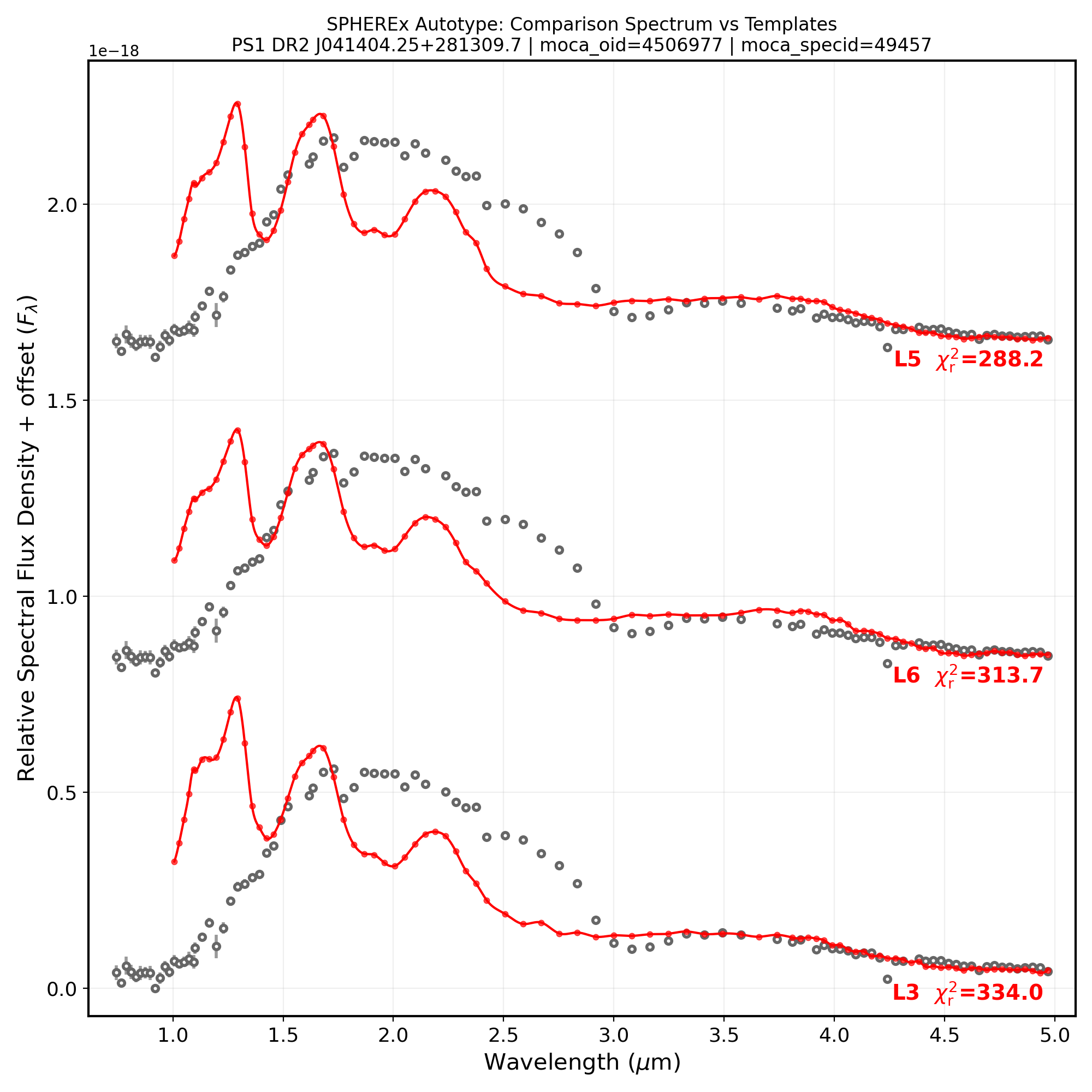}\label{fig:autotypebcg}}
    \subfigure[An ambiguous case due to low-quality data]{\includegraphics[width=0.47\textwidth]{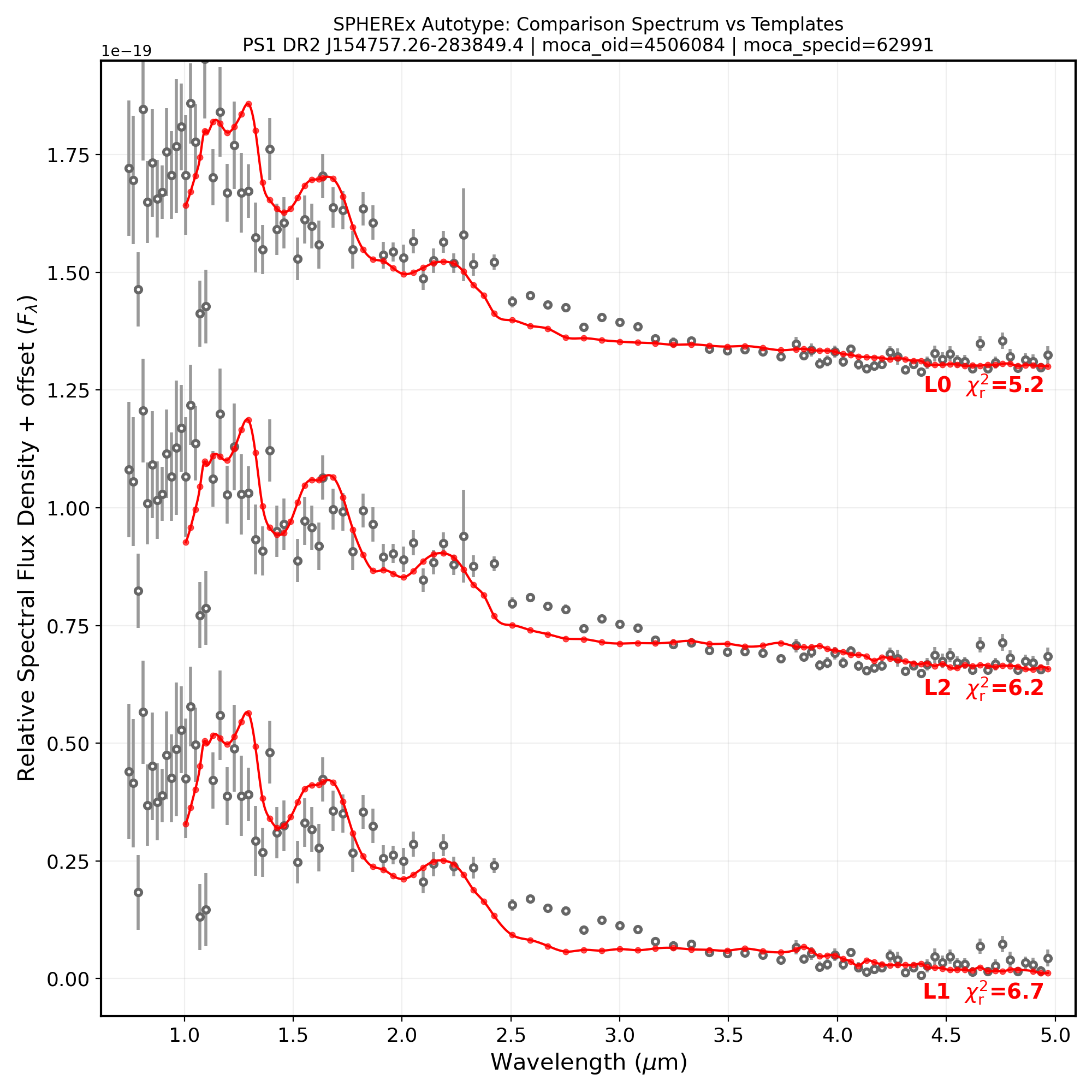}\label{fig:autotypeamb}}
    \subfigure[Confusion between two close sources]{\includegraphics[width=0.47\textwidth]{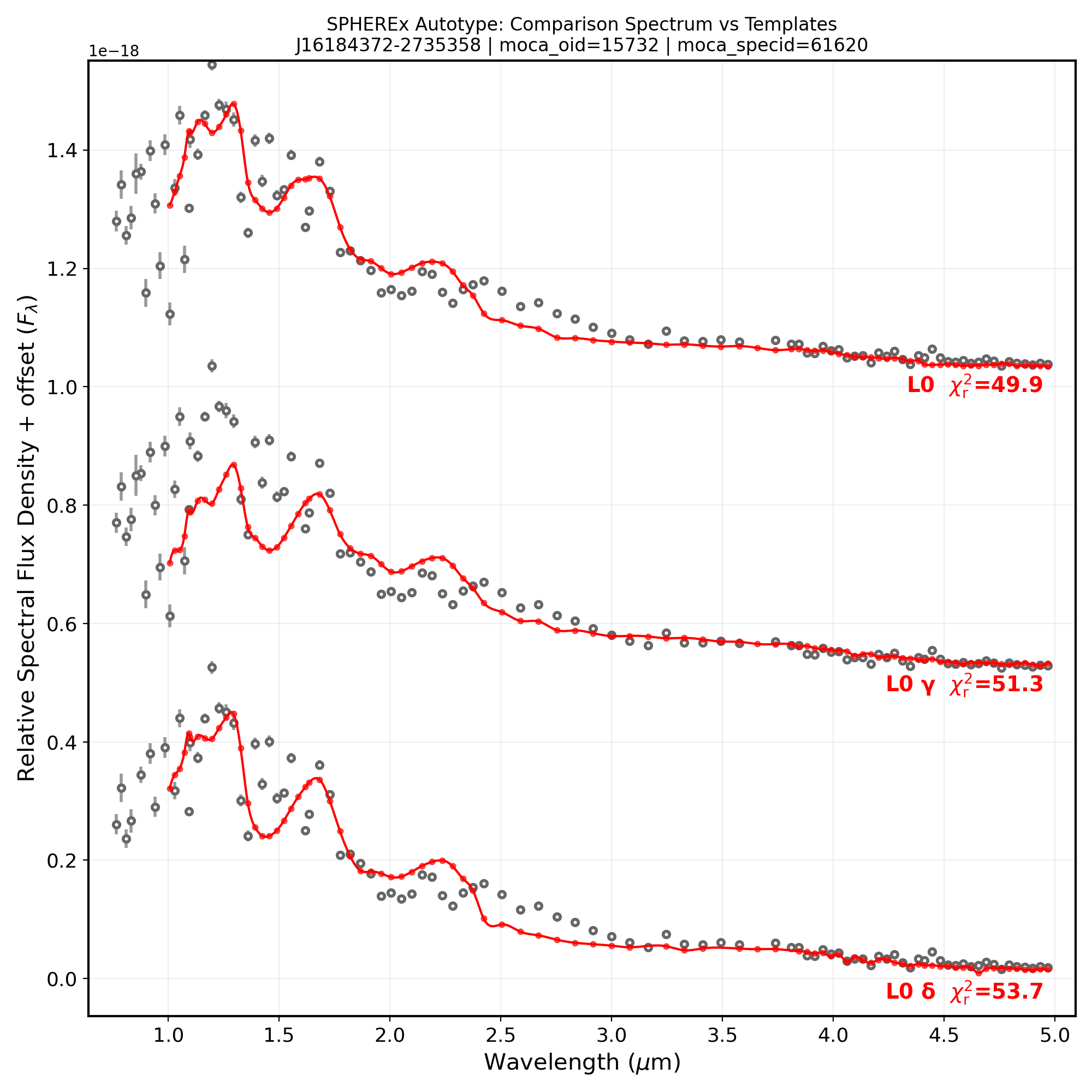}\label{fig:autotypeconfusion}}
	\caption{The four common failure modes resulting in poor-quality SPIFF spectrophotometry. Upper left: an unresolved contaminating background star dominates the signal (occurred for \percfailurestar\%\ of known \added{ultracool} dwarfs). Upper right: A false-positive brown dwarf corresponding to a reddened background star (occurred for \percfailurereddened\%\ of known \added{ultracool} dwarfs). Lower left: An ambiguous spectrum due to low S/N data (occurred for \percfailuresnr\%\ of known \added{ultracool} dwarfs with W2$\leq$\wmagsnr\,mag). Lower right: Two or more resolved but closely separating sources causing confusion, where the PSF model can converge to either source depending on the epoch (occurred for \percfailurescatter\%\ of known \added{ultracool} dwarfs). See Section~\ref{sec:conf_cand} for more details.}
	\label{fig:failures}
\end{figure*}

Automated spectral types were determined using $\chi^2$ fitting of each \added{hybrid} spectral template applied to every spectrum of the SPIFF library. When doing this, the worst 5 data points are always ignored to make spectral typing more robust against outlier data points, and the optimal scaling factor that minimizes $\chi^2$ is determined analytically, independently for each \added{hybrid} template. Additionnally, non-field \added{hybrid} templates (young brown dwarfs or subdwarfs) are assigned a penalty in their total $\chi^2$ score of \smallpenalty, and extremely low-gravity or extreme subdwarfs are assigned a larger penalty of \largepenalty, to avoid classifying low-S/N spectra as potential subdwarfs or young objects. Figures comparing the SPIFF spectrophotometry to the best three \added{hybrid} templates were generated for each literature \added{ultracool} dwarf and were then visually vetted to flag contaminated or otherwise bad-quality spectra. The high-quality spectra are shown in Figure~\ref{fig:spt_fitting}, and the four typical failure modes or resulting bad-quality spectrophotometry are shown in Figure~\ref{fig:failures}.

\begin{figure}
	\centering
	\includegraphics[width=\linewidth]{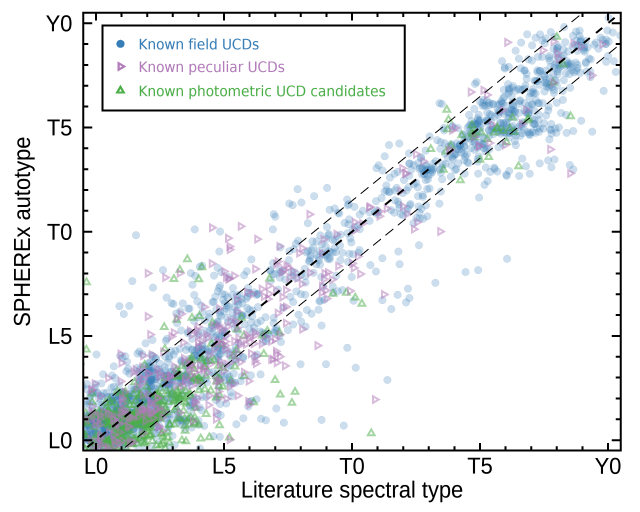}
	\caption{Validation of the automated SPHEREx spectral types compared with literature spectral types for known ultracool dwarfs (blue circles or field objects, rightward purple triangles for known peculiar UCDs), and for photometric ultracool dwarf candidates with spectral type estimates in the literature (upward green triangles). The thick, dashed line represents the 1:1 relation and the thinner dashed lines represent the median absolute deviation (\autotypemad\ subtypes). A random jitter of 0.3 subtypes was added to both spectral type dimensions to improve visibility. See Section~\ref{sec:spt} for more details.}
	\label{fig:sptverif}
\end{figure}

\begin{figure*}[!t]
	\centering
    \subfigure[Spectral type versus \added{reduced} $\chi^2$]{\includegraphics[width=0.48\textwidth]{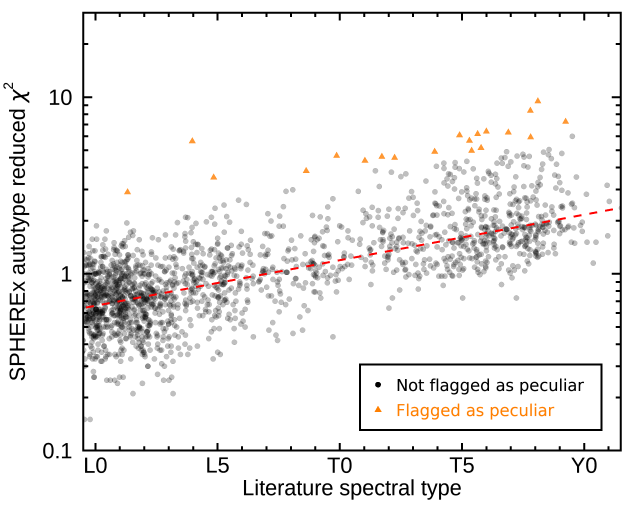}\label{fig:chi2_spt}}
    \subfigure[WISE W2 magnitude versus \added{reduced} $\chi^2$]{\includegraphics[width=0.48\textwidth]{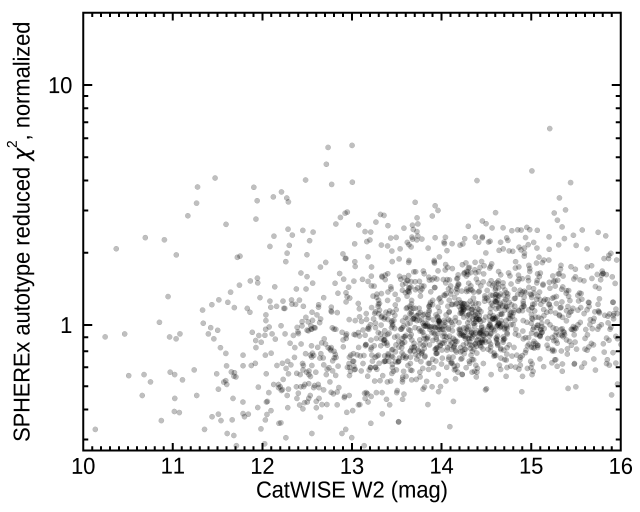}\label{fig:chi2_w2}}
	\caption{Autotyping reduced $\chi^2$ as a function of spectral type (left) and W2 magnitude (right). We included a floor of 10\% on the minimum SPIFF error bars to compute this $\chi^2$ (as a guard against high-S/N outliers) and adjust a robust linear fit as a function of spectral type (red line in the left panel) to compensate for the systematic trend observed here. All objects with $\chi^2 > $ \chilimit\ were flagged as potentially peculiar targets. See Section~\ref{sec:conf_cand} for more details.}
	\label{fig:chi2}
\end{figure*}

Figure~\ref{fig:sptverif} shows the comparison in automated spectral types with literature spectral types, showing good agreement with a median absolute deviation of \autotypemad\ subtypes. We caution that SPIFF data products that are contaminated by background stars can be significantly more discrepant. Figure~\ref{fig:chi2} shows the fitting reduced $\chi^2$ as a function of spectral types and W2 magnitudes.

All template-fitting figures were inspected visually to assess the quality of the resulting SPHEREx spectrophotometry data products output by SPIFF. We used the set of known \added{ultracool} dwarfs with previous spectroscopic confirmation and a WISE W2 magnitude \wmagsnr\ or brighter as a test sample to establish the different failure modes and their respective occurrence rates. Out of 1327 known \textbf{brown} dwarfs with previous spectroscopic confirmation that have W2 $\leq$ 14.0, we classified 1078 (\percgood\%) has having high-quality SPHEREx spectrophotometry clearly matching an ultracool \added{hybrid} template, 114 (\percfailurestar\%) as having a slightly earlier spectral type (M8--M9), 14 (\percfailbdplusstar\%) as a blended \added{ultracool} dwarf and background star spectra, 20 (\percamb\%) as ambiguous cases, and 101 (\percbad\%) as a low-quality spectrum. Out of the 101 bad cases, 4 (\percfailurereddened\%) were classified as background, reddened sources, 4 (\percfailuresnr\%) were classified as having a S/N too low for proper classification, 35 (\percfailurescatter\%) were classified as a clear mismatch where the signal is dominated by a background star with a much earlier spectral type, and 58 (\percfailurescatter\%) as blended spectra from more than one source. We note that these problematic cases likely indicate a contaminated SPIFF data product, and not necessarily a rejection of the ultracool status of the targets, given that contamination from background stars can play an important role with the large pixel size of SPHEREx.

\begin{figure}
	\centering
	\includegraphics[width=\linewidth]{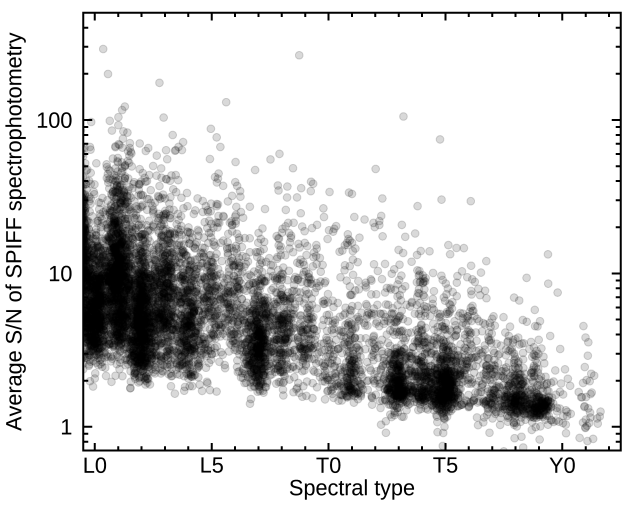}
	\caption{Average signal-to-noise ratio of the spectrophotometry products compiled here as a function of spectral type. A random jitter of 0.3 subtypes was added for clarity. See Section~\ref{sec:spt} for more details.}
	\label{fig:snrlib}
\end{figure}

The final library of SPIFF spectrophotometry products that were visually vetted cover a total of \nhqspectra\ UCDs. Their average signal-to-noise ratio as a function of spectral type are presented in Figure~\ref{fig:snrlib}.

\subsection{Confirmation of Ultracool Dwarf Candidates}\label{sec:conf_cand}

We constructed SPHEREx spectrophotometry for a set of \added{ultracool} dwarf candidates described in Section~\ref{sec:sample}, applied the automated spectral typing method of Section~\ref{sec:spt}, and visually inspected all resulting template-fitting figures to confirm the nature of \added{ultracool dwarfs} in an attempt to demonstrate the power of SPHEREx in distinguishing photometrically selected \added{ultracool} dwarfs from contaminants such as reddened background stars and galaxies, especially in the absence of reliable proper motion detection.

\begin{figure}
	\centering
	\includegraphics[width=\linewidth]{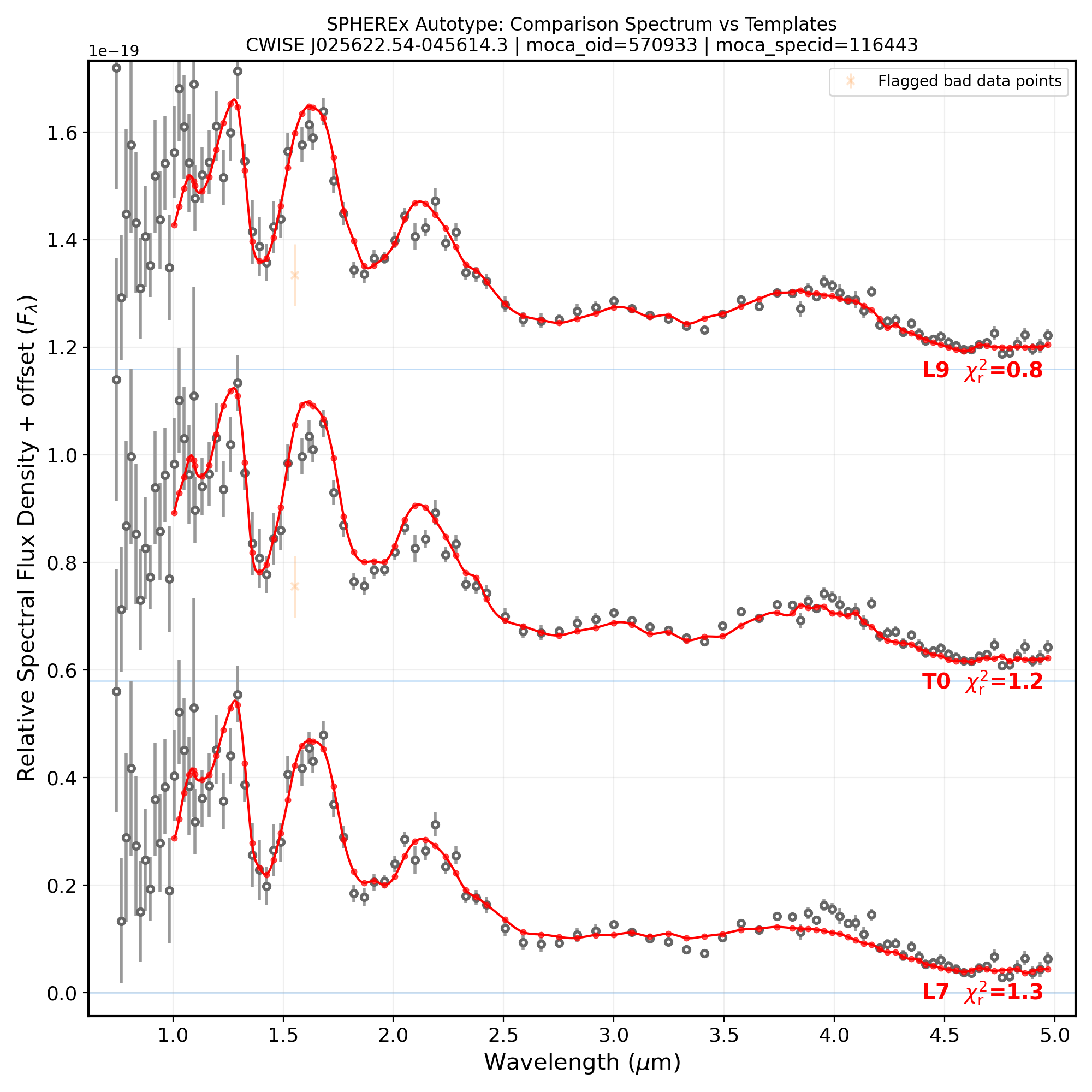}
	\caption{Confirmation of new and known ultracool dwarf candidates. The complete figure set (\nhqcand\ images) is available in the online journal. See Section~\ref{sec:conf_cand} for more details.}
	\label{fig:newcands}
\end{figure}

We visually verified the \added{hybrid} template matches for all SPHEREx spectrophotometry with flux measurements available in at least \datapointslimit/102 of the SPHEREx channels. This allowed us to \nhqcand\ \added{ultracool dwarfs} with spectral types L0--Y1, \nbyw\ of which are from the Backyard Worlds set of high-proper motion candidates \citep{2026arXiv260401323S}. \added{This represents more than a doubling of the number of ultracool dwarfs with spectroscopic confirmation.} All newly confirmed \added{ultracool dwarfs} are shown in Figure~\ref{fig:newcands}.

There are \ncanducd\ low-S/N SPHEREx data products that \added{we classify as} \added{`New UCD candidates'} and hint at water, carbon dioxide and/or methane features but that we do not consider of high-enough quality to warrant a confirmation of the UCD nature of the target (see Figure~\ref{fig:nspectra_spt}). We categorize these cases as new UCD candidates, and further follow-up with higher-S/N spectroscopy will be required to confirm them.

\subsection{Peculiar Spectra}\label{sec:pec}

\begin{figure}
	\centering
	\includegraphics[width=\linewidth]{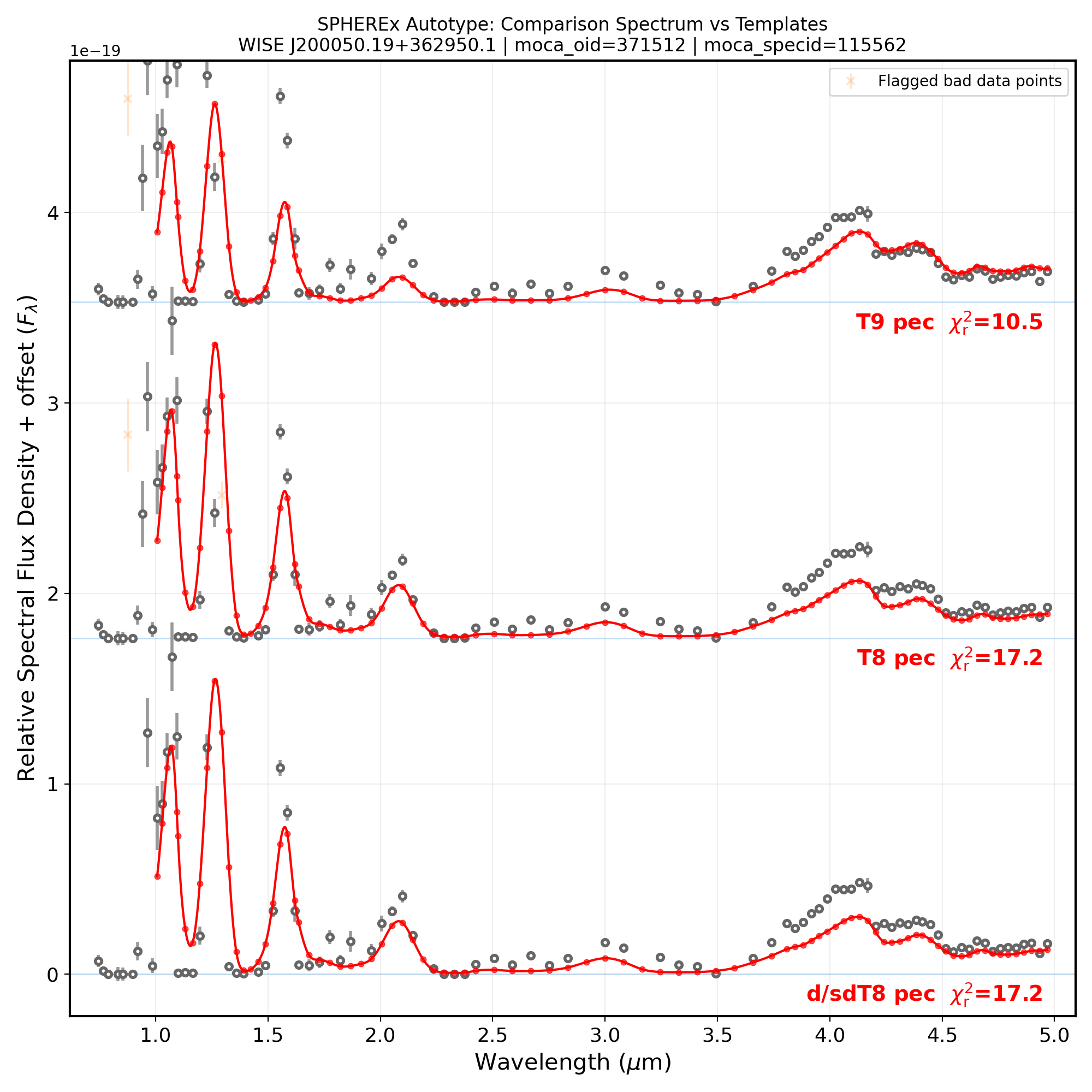}
	\caption{Peculiar spectra identified by a visual comparison of best-matching \added{hybrid} templates with high-quality SPHEREx spectra. The example object shown is a known T8 \citep{2014AJ....147..113C} with an extreme $\approx$4.2$\mu$m CO$_2$ feature for its spectral type. The complete figure set (\nspecial\ images) is available in the online journal. See Section~\ref{sec:pec} for more details.}
	\label{fig:pec}
\end{figure}

\added{Based on our inspection of initial SPIFF classifications of hybrid templates, we flagged a set of  \nspecial\ spectra with potentially peculiar features, because they were best-matched to a subdwarf or young hybrid template, or because of the high $\chi^2$ of the best hybrid template match.} These fall in three categories: low-gravity brown dwarfs (\nknyoung\ previously known young brown dwarfs, \nnewyoung\ new young candidates, although some in the later category may have implicitly been called young in the literature from their membership in young associations); subdwarfs (\nknsd\ previously known subdwarfs, \nnewsd\ new subdwarf candidates); and otherwise peculiar spectra (\nknpec\ previously flagged as peculiar, and \nnewpec\ newly flagged as such). These peculiar spectra are shown in Figure~\ref{fig:pec} and listed in Table~\ref{tab:spt}\added{ sorted by dataset and from late to early spectral types, and would be interesting candidates for future spectroscopic characterization}. We caution that some subdwarf \added{hybrid} templates (e.g., esdL8, sdT4) resemble a blend between a T dwarf and an early-type star across 1--5\,$\mu$m, and therefore these \added{hybrid} templates are often attributed to cases with a background star contaminating a SPHEREx pixel. These cases should therefore be treated with care; we generally adopted only those with a high S/N and clear detections of water or methane that were not within 10\arcsec\ of a background star.

\subsection{Access to Data Products}\label{sec:datap}

We provide \added{both} our final set of SPHEREx \added{hybrid and raw} templates in the online Data Behind Figures product of Figures \ref{fig:l_templates} and \ref{fig:ty_templates}, along with the current snapshot of high-quality SPHEREx spectra, both as a package of \texttt{csv} spectra, and as template-matching figures in the online-only material associated with this work. All data products are also available on the MOCA database\footnote{Available at \url{https://mocadb.ca}.} \citep{2024PASP..136f3001G,2026arXiv260215695G}, and will be updated there, as well as in a persistent and versioned Zenodo repository at \href{https://doi.org/10.5281/zenodo.19051216}{doi:10.5281/zenodo.19051216}, as new SPHEREx data products become available.

\section{CONCLUSIONS}\label{sec:conclusion}

We present a new data reduction package for SPHEREx spectrophotometry, designed to recover high-quality data sets for fast-moving objects such as nearby brown dwarfs. We compiled a set of \nhqspectra\ high-quality 0.75--5.0\,$\mu$m spectra for \nhqknown\ known \added{ultracool} dwarfs and \nhqcand\ newly confirmed ones, along with tools to automatically determine their spectral types based on \added{hybrid} spectral templates. These data sets \added{already more than double the number of ultracool dwarfs with spectroscopic confirmation and} will be refined further as the SPHEREx mission progresses and will facilitate several follow-up studies benefiting from both the wavelength and temporal coverages provided by SPHEREx. Such examples include: the empirical determination of bolometric luminosities for a large number of brown dwarfs; the search for unresolved Y dwarf companions; detailed atmospheric properties across a wide range of atmospheric layers with spectral retrieval methods; brown dwarf variability studies; the identification of brown dwarfs with peculiar properties (e.g., subdwarfs, young brown dwarfs or chemically peculiar objects); the search or confirmation of protoplanetary disks around brown dwarfs; and the spectroscopic confirmation of new \added{ultracool} dwarf candidates.

\begin{acknowledgements}

\added{We thank the anonymous reviewer for their helpful comments that improved the quality of this work.} We would like to thank the IPAC helpdesk for their generous help with this project. We thank Jean-Maxime Couillard for his technical support with this project. J.G. and A.R.D. acknowledge the support of the Canadian Space Agency (CSA) [\textbf{25JWGO4B10}], and J.G. acknowledges the support of the Natural Sciences and Engineering Research Council of Canada (NSERC), funding reference number RGPIN-2021-03121. L.-P.C. acknowledges support from Mitacs through the Mitacs Accelerate program, in partnership with the Montreal Planetarium.  J.F acknowledges the support of NSF CAREER award number 2238468.

This research was enabled in part by support provided by Universit\'e de Montr\'eal (\url{umontreal.ca}) and the Digital Research Alliance of Canada (\url{alliancecan.ca}).
This publication makes use of data products from the Spectro-Photometer for the History of the Universe, Epoch of Reionization and Ices Explorer (SPHEREx), which is a joint project of the Jet Propulsion Laboratory and the California Institute of Technology, and is funded by the National Aeronautics and Space Administration. This paper uses the SPHEREx Quick Release Spectral Images -- QR2 \dataset[10.26131/IRSA652]{http://dx.doi.org/10.26131/IRSA652}.

This research made use of the Montreal Open Clusters and Associations (MOCA) database, operated at the Montr\'eal Plan\'etarium \citep{2024PASP..136f3001G,2026arXiv260215695G}. This work has benefited from The UltracoolSheet \citep{Best20US} at \url{http://bit.ly/UltracoolSheet}, maintained by Will Best, Trent Dupuy, Michael Liu, Rob Siverd, and Zhoujian Zhang, and developed from compilations by \cite{2012ApJS..201...19D}, \cite{2013Sci...341.1492D}, \cite{2016ApJ...833...96L}, \cite{2018ApJS..234....1B}, and \cite{2021AJ....161...42B}. This work is based in part on observations made with the NASA/ESA/CSA James Webb Space Telescope. The data were obtained from the Mikulski Archive for Space Telescopes at the Space Telescope Science Institute, which is operated by the Association of Universities for Research in Astronomy, Inc., under NASA contract NAS 5-03127 for JWST. These observations are associated with programs GO~3548, GO~2302, GTO~1189, and GTO~4539.

This research made use of: the SIMBAD database and VizieR catalog access tool, operated at the Centre de Donn\'ees astronomiques de Strasbourg, France \citep{2000AAS..143....9W}; data products from the Two Micron All Sky Survey (\emph{2MASS}; \citealp{2006AJ....131.1163S}), which is a joint project of the University of Massachusetts and the Infrared Processing and Analysis Center (IPAC)/California Institute of Technology (Caltech), funded by the National Aeronautics and Space Administration (NASA) and the National Science Foundation \citep{2006AJ....131.1163S}; data products from the \emph{Wide-field Infrared Survey Explorer} (\emph{WISE}; and \citealp{2010AJ....140.1868W}), which is a joint project of the University of California, Los Angeles, and the Jet Propulsion Laboratory (JPL)/Caltech, funded by NASA. This work has made use of the SIMPLE Archive of low-mass stars, brown dwarfs, and directly imaged exoplanets: \texttt{10.5281/zenodo.13937301}.

All the {\it JWST} data used in this paper can be found in MAST: \dataset[10.17909/edj9-e737]{http://dx.doi.org/10.17909/edj9-e737}, \dataset[10.17909/4wg9-8196]{http://dx.doi.org/10.17909/4wg9-8196}.

\end{acknowledgements}

\bibliographystyle{aasjournalv7}
\bibliography{cited_refs}
\newpage
\appendix

In this Appendix, we provide example MySQL queries which can be used in the MOCAdb to retrieve distinct SPHEREx SPIFF data products.

The main data products are spectrophotometry in $F_\lambda$ binned over the 102 SPHEREx spectral channels, and can be acquired for a single object using:
\lstset{basicstyle=\ttfamily, tabsize=4}
\begin{lstlisting}[language=SQL, frame=single]
-- Obtain one SPHEREx data product for SIMP J0136
--   (binned per 102 SPHEREx channel)
SELECT
	-- Convert wavelengths to microns
	ds.wavelength_angstrom/1e4 AS wv_microns,
	-- Convert absolute F_lambda (W/m^2/Angstroms) to F_lambda (W/m^2/microns)
	ds.flux_flambda/1e4 AS flux,
	-- Same with errors
	ds.flux_flambda_unc/1e4 AS flux_err
-- Grab the MOCAdb object ID based on designation
FROM mechanics_all_designations mad
-- Grab spectra headers by matching unique MOCAdb object ID
JOIN moca_spectra ms USING(moca_oid)
-- Grab spectra data points by matching unique MOCAdb spectrum ID
JOIN data_spectra ds USING(moca_specid)
WHERE 
	-- Specify object designation
	mad.designation='SIMP J013656.5+093347.3'
	-- Only select latest (ignored=0) SPHEREx SPIFF
    --   reduction (moca_specpackid=55).
    --   Use moca_specpackid=54 for the (rare) IRSA-pipeline
    --   SPHEREx data products stored in MOCAdb
	AND ms.ignored=0 AND ms.moca_specpackid=55;
\end{lstlisting}

The original SPIFF outputs based on \texttt{ultranest} samples in $F_\nu$, often with repeated wavelengths and individual observing epochs, can be obtained with:
\begin{lstlisting}[language=SQL, frame=single]
-- Obtain one SPHEREx data product for SIMP J0136
--   (time-resolved, not binned per channel, based on ultranest)
SELECT
	-- Grab the exact bandpass wavelength center and width
	--   at that specific detector location
	ds.psf_un_wv_um AS wv_um, ds.psf_un_wv_width_um AS wv_wid_um, 
	-- Grab F_nu flux (microjanskies) and error
	ds.psf_un_flux_uJy AS flux, ds.psf_un_flux_uJy_err AS flux_err,
	-- Grab the exact MJD epoch of this SPHEREx FITS
	--   image (useful for time-resolved analysis)
	ds.mjd_avg
-- Grab the MOCAdb object ID based on designation
FROM mechanics_all_designations mad
-- Grab the raw SPIFF data points by matching unique MOCAdb object ID
JOIN data_spherex_spectra_spiff ds USING(moca_oid)
WHERE 
	-- Specify object designation
	mad.designation='SIMP J013656.5+093347.3'
	-- Reject any data points flagged as bad
	AND ds.ignored=0;
\end{lstlisting}
\newpage
Spectrophotometry based on the \texttt{scipy} optimizer without \texttt{ultranest} sampling can be obtained with:
\begin{lstlisting}[language=SQL, frame=single]
-- Obtain one SPHEREx data product for SIMP J0136
--   (time-resolved, not binned per channel, based on scipy)
SELECT
	-- Grab the exact bandpass wavelength center and width
	--   at that specific detector location
	ds.psf_scipy_wv_um AS wv_um, ds.psf_scipy_wv_width_um AS wv_wid_um, 
	-- Grab F_nu flux (microjanskies) and error
	ds.psf_scipy_flux_uJy AS flux, ds.psf_scipy_flux_uJy_err AS flux_err,
	-- Grab the exact MJD epoch of this SPHEREx FITS
	--   image (useful for time-resolved analysis)
	ds.mjd_avg
-- Grab the MOCAdb object ID based on designation
FROM mechanics_all_designations mad
-- Grab the raw SPIFF data points by matching unique MOCAdb object ID
JOIN data_spherex_spectra_spiff ds USING(moca_oid)
WHERE 
	-- Specify object designation
	mad.designation='SIMP J013656.5+093347.3'
	-- Reject any data points flagged as bad
	AND ds.ignored=0;
\end{lstlisting}

Spectrophotometry based on aperture photometry (which requires extreme care for background contamination) can be obtained with:
\begin{lstlisting}[language=SQL, frame=single]
-- Obtain one SPHEREx data product for SIMP J0136
--   (time-resolved, not binned per channel, based on aperture photometry)
SELECT
	-- Grab the exact bandpass wavelength center and width
	--   at the detector location corresponding to the
    --   photometric center-of-mass
	ds.com_wv_um AS wv_um, ds.com_wv_width_um AS wv_wid_um, 
	-- Grab F_nu flux (microjanskies) and error
	ds.ap_flux_uJy AS flux, ds.ap_flux_uJy_err AS flux_err,
	-- Grab the exact MJD epoch of this SPHEREx FITS
	--   image (useful for time-resolved analysis)
	ds.mjd_avg
-- Grab the MOCAdb object ID based on designation
FROM mechanics_all_designations mad
-- Grab the raw SPIFF data points by matching unique MOCAdb object ID
JOIN data_spherex_spectra_spiff ds USING(moca_oid)
WHERE 
	-- Specify object designation
	mad.designation='SIMP J013656.5+093347.3'
	-- Reject any data points flagged as bad
	AND ds.ignored=0;
\end{lstlisting}

\input{tables/data_spherex_spectra_spiff_preview}
\input{tables/pcat_spherex_bins_bandpasses_preview}
\input{tables/spherex_template_grids_fulltable}
\newpage
\input{tables/spherex_autotype_datasets_preview}

\end{document}

%% file: tables/data_spherex_spectra_spiff_preview.tex
\startlongtable
\begin{deluxetable*}{lll}
\tablecaption{Preview of columns in \texttt{data\_spherex\_spectra\_spiff}}
\label{tab:spiff}
\tablehead{
\colhead{Column} & \colhead{Datatype} & \colhead{Description}
}
\startdata
id & bigint(20) unsigned & Primary identifier for this row. \\
md5\_uid & varchar(32) & Deterministic MD5 UID built from pipeline versio(...) \\
moca\_oid & int(10) unsigned & MOCA object identifier (foreign key to moca\_obje(...) \\
moca\_specid & int(10) unsigned & MOCA spectrum identifier linked to this SPIFF ro(...) \\
target\_name & varchar(255) & Human-readable target name passed to LV2 for thi(...) \\
reference\_ra\_deg & double & Reference right ascension in degrees used for PM(...) \\
reference\_dec\_deg & double & Reference declination in degrees used for PM pro(...) \\
reference\_crd\_epoch\_yr & double & Epoch (year) of the reference sky coordinates. \\
reference\_pmra\_masyr & double & Reference proper motion in RA*cos(Dec), in mas/yr. \\
reference\_pmdec\_masyr & double & Reference proper motion in Dec, in mas/yr. \\
input\_ra\_deg & double & RA in degrees actually sent to the image analyze(...) \\
input\_dec\_deg & double & Dec in degrees actually sent to the image analyz(...) \\
\enddata
\tablecomments{This table is available in its entirety online. The full table contains 112 rows.}
\end{deluxetable*}

%% file: tables/pcat_spherex_bins_bandpasses_preview.tex
\startlongtable
\begin{deluxetable}{cll}
\tablewidth{\textwidth}
\tablecaption{Preliminary SPHEREx channel characteristics (Preview)\label{tab:bandpass}}
\tablehead{
\colhead{Index} & \colhead{$\lambda_{\mathrm{center}}$ ($\mu$m)} & \colhead{$\lambda_{\mathrm{width}}$ ($\mu$m)}
}
\startdata
1 & 0.744 & 0.022 \\
2 & 0.766 & 0.022 \\
3 & 0.788 & 0.022 \\
4 & 0.810 & 0.022 \\
5 & 0.832 & 0.022 \\
6 & 0.853 & 0.022 \\
7 & 0.875 & 0.022 \\
8 & 0.897 & 0.022 \\
9 & 0.919 & 0.022 \\
10 & 0.941 & 0.022 \\
11 & 0.963 & 0.022 \\
12 & 0.985 & 0.022 \\
\enddata
\tablecomments{This table is available in its entirety online. The full table contains 102 rows.}
\end{deluxetable}

%% file: tables/spherex_template_grids_fulltable.tex
\startlongtable
\begin{deluxetable*}{lllrll}
\tablecaption{SPHEREx Template Grids (Full Table)}
\label{tab:std}
\tablehead{
\colhead{Grid Type} & \colhead{Spectral} & \colhead{Instrument} & \colhead{$N_{\mathrm{spec}}$\tablenotemark{a}} & \colhead{Standard Designation\tablenotemark{b}} & \colhead{Ref.} \\
\colhead{} & \colhead{Type} & \colhead{} & \colhead{} & \colhead{} & \colhead{}
}
\startdata
Field & M8 & SpeX IRTF & 76 & VB 10 & 1 \\
Field & M9 & SpeX IRTF & 71 & LHS 2924 & 1 \\
Field & L0 & SpeX IRTF & 544 & 2MASP J0345432+254023 & 1 \\
Field & L1 & SpeX IRTF & 399 & 2MASS J21304464-0845205 & 1 \\
Field & L2 & SpeX IRTF & 382 & 2MASSI J0408290-145033 & 2 \\
Field & L3 & SpeX IRTF & 178 & 2MASSW J1506544+132106 & 3 \\
Field & L4 & SpeX IRTF & 132 & 2MASS J21580457-1550098 & 1 \\
Field & L5 & SpeX IRTF & 112 & 2MASSI J0652307+471034 & 3 \\
Field & L6 & SpeX IRTF & 56 & 2MASSI J1010148-040649 & 1 \\
Field & L7 & SpeX IRTF & 57 & 2MASS J02052940-1159296 & 4 \\
Field & L8 & SpeX IRTF & 35 & 2MASSW J1632291+190441 & 1 \\
Field & L9 & SpeX IRTF & 51 & 2MASS J02550357-4700509 & 1 \\
Field & T0 & SpeX IRTF & 31 & SIMP J0150+3827 & 5 \\
Field & T1 & Nires Keck & 38 & SDSS J083717.21-000018.0 & 1 \\
Field & T2 & SpeX IRTF & 56 & SDSS J125453.90-012247.4 & 1 \\
Field & T3 & SpeX IRTF & 47 & 2MASS J12095613-1004008 & 6 \\
Field & T4 & SpeX IRTF & 58 & 2MASSI J2254188+312349 & 6 \\
Field & T5 & SpeX IRTF & 85 & 2MASS J15031961+2525196 & 6 \\
Field & T6 & SpeX IRTF & 0 & SDSSp J162414.37+002915.6 & 6 \\
Field & T7 & SpeX IRTF & 0 & 2MASSI J0727182+171001 & 6 \\
Field & T8 & SpeX IRTF & 0 & 2MASSI J0415195-093506 & 6 \\
Field & d/sdT8 & SpeX IRTF & 0 & 2MASSI J0415195-093506 & 6 \\
Field & T9 & Nires Keck & 0 & UGPS J072227.51-054031.2 & 6 \\
Field & Y0 & NIRSpec JWST & 0 & WISEA J120604.25+840110.5 & 7 \\
Field & Y1 & NIRSpec JWST & 0 & WISE J053516.80-750024.9 & 8 \\
Intermediate gravity & M8 $\beta$ & SpeX IRTF & 22 & 2MASSI J2323134-024435 & 9 \\
Intermediate gravity & M9 $\beta$ & SpeX IRTF & 7 & 2MASSI J1411213-211950 & 10 \\
Intermediate gravity & L0 $\beta$ & SpeX IRTF & 15 & 2MASS J11544223-3400390 & 2 \\
Intermediate gravity & L1 $\beta$ & SpeX IRTF & 22 & 2MASS J10224821+5825453 & 2 \\
Intermediate gravity & L2 $\beta$ & SpeX IRTF & 19 & LSPM J0602+3910 & 11 \\
Intermediate gravity & L3 $\beta$ & SpeX IRTF & 19 & 2MASS J22495345+0044046 & 11 \\
Intermediate gravity & L4 $\beta$ & SpeX IRTF & 11 & 2MASSW J0030300-145033 & 9 \\
Very low gravity & M9 $\gamma$ & SpeX IRTF & 6 & TWA 28 & 9 \\
Very low gravity & L0 $\gamma$ & SpeX IRTF & 25 & 2MASS J01415823-4633574 & 2 \\
Very low gravity & L1 $\gamma$ & SpeX IRTF & 21 & 2MASSI J0518461-275645 & 2 \\
Very low gravity & L2 $\gamma$ & SpeX IRTF & 9 & 2MASSI J0536199-192039 & 2 \\
Very low gravity & L3 $\gamma$ & Flamingos-2 Gemini South & 13 & 2MASS J04185879-4507413 & 9 \\
Very low gravity & L4 $\gamma$ & SpeX IRTF & 16 & 2MASS J16154255+4953211 & 2 \\
Very low gravity & L5 $\beta$/$\gamma$ & SpeX IRTF & 9 & 2MASS J03264225-2102057 & 9 \\
Very low gravity & L7 $\gamma$ & SpeX IRTF & 8 & PSO J318.5338-22.8603 & 12 \\
Extremely low gravity & L0 $\delta$ & SpeX IRTF & 7 & 2MASS J06085283-2753583 & 9 \\
Slight subdwarfs & d/sdL8 & SpeX IRTF & 28 & 2MASS J11582077+0435014 & 6 \\
Slight subdwarfs & d/sdT1 & Fire Magellan & 2 & WISEA J030119.39-231921.1 & 6 \\
Subdwarfs & sdM9.5 & SpeX IRTF & 4 & 2MASS J10130734-1356204 & 13 \\
Subdwarfs & sdL3.5 & SpeX IRTF & 4 & 2MASS J12563716-0224522 & 1 \\
Subdwarfs & sdT4 & Nires Keck & 3 & WISE J155349.98+693355.2 & 6 \\
Extreme subdwarfs & esdL8 & Nires Keck & 2 & 2MASS J05325346+8246465 & 6 \\
\enddata
\tablenotetext{a}{Number of SPHEREx spectra used for the construction of this template.}
\tablenotetext{b}{Designation of the object we used as a spectral standard for wavelengths below 2.4\,$\mu$m. References specify which work established their spectral type.}
\tablecomments{References: (1) \cite{2010ApJS..190..100K}; (2) \cite{2018AJ....155...34C}; (3) \cite{2009AJ....137.3345C}; (4) \cite{1999ApJ...519..802K}; (5) \cite{2023AJ....166..226B}; (6) \cite{2025ApJ...982...79B}; (7) \cite{2015ApJ...804...92S}; (8) \cite{2012ApJ...753..156K}; (9) \cite{2015ApJS..219...33G}; (10) \cite{2003AJ....126.2421C}; (11) \cite{2013ApJ...772...79A}; (12) \cite{2013ApJ...777L..20L}; (13) \cite{2019AA...628A..61L}.}
\end{deluxetable*}

%% file: tables/spherex_autotype_datasets_preview.tex
\startlongtable
\begin{deluxetable*}{lllccccc}
\tabletypesize{\scriptsize}
\tablecaption{SPHEREx Spectral Classifications for High-Quality SPIFF Spectrophotometry Products (Preview)}
\label{tab:spt}
\tablehead{
\colhead{Designation} & \colhead{RA} & \colhead{Dec} & \colhead{Dataset} & \colhead{SPIFF} & \colhead{$\chi^2_{\mathrm{r}}$} & \colhead{Literature} & \colhead{Ref.} \\
\colhead{} & \colhead{J2000} & \colhead{J2000} & \colhead{} & \colhead{Autotypes} & \colhead{} & \colhead{Spectral Types} & \colhead{}
}
\startdata
CWISE J101438.37-272622.7 & 10:14:38.37 & -27:26:22.8 & New Discoveries & T9 & 1.6 & \nodata & \nodata \\
CWISE J232834.16+183438.1 & 23:28:34.17 & +18:34:38.1 & New Discoveries & T9 & 1.1 & \nodata & \nodata \\
CWISE J233232.63+232559.0 & 23:32:32.63 & +23:25:59.0 & New Discoveries & T9 & 1.8 & \nodata & \nodata \\
UHS DR3 459624484576 & 01:51:26.04 & +42:27:03.4 & New Discoveries & T9 & 1.3 & \nodata & \nodata \\
UHS DR3 459702022403 & 23:28:46.63 & +31:39:17.4 & New Discoveries & T9 & 1.4 & \nodata & \nodata \\
UHS DR3 459758596359 & 22:40:51.02 & +36:25:31.4 & New Discoveries & T9 & 1.0 & \nodata & \nodata \\
UHS DR3 459761069871 & 00:52:58.95 & +47:56:06.0 & New Discoveries & T9 & 1.0 & \nodata & \nodata \\
UHS DR3 460104839527 & 23:38:02.12 & +38:53:04.9 & New Discoveries & T9 & 1.3 & \nodata & \nodata \\
CWISE J043424.24+093228.3 & 04:34:24.24 & +09:32:28.5 & New Discoveries & T8 & 1.3 & \nodata & \nodata \\
CWISE J055506.02+394701.6 & 05:55:06.03 & +39:47:01.7 & New Discoveries & T8 & 1.5 & \nodata & \nodata \\
CWISE J080858.88-125605.6 & 08:08:58.89 & -12:56:05.6 & New Discoveries & T8 & 1.3 & \nodata & \nodata \\
CWISE J094449.87+265451.6 & 09:44:49.87 & +26:54:51.6 & New Discoveries & T8 & 1.6 & \nodata & \nodata \\
CWISE J094615.56+351434.0 & 09:46:15.56 & +35:14:34.0 & New Discoveries & T8 & 1.7 & \nodata & \nodata \\
CWISE J010134.74+403832.8 & 01:01:34.74 & +40:38:32.8 & New Discoveries & T8 & 1.4 & \nodata & \nodata \\
CWISE J221355.02-392301.2 & 22:13:55.02 & -39:23:01.2 & New Discoveries & T8 & 1.5 & \nodata & \nodata \\
UHS DR3 459568471419 & 17:37:07.04 & +44:58:46.0 & New Discoveries & T8 & 1.9 & \nodata & \nodata \\
UHS DR3 459659536509 & 23:12:41.24 & +39:26:53.4 & New Discoveries & T8 & 1.8 & \nodata & \nodata \\
UHS DR3 459687123729 & 02:24:20.41 & +36:33:53.0 & New Discoveries & T8 & 1.4 & \nodata & \nodata \\
UHS DR3 459775185717 & 07:59:31.48 & +57:59:25.5 & New Discoveries & T8 & 1.2 & \nodata & \nodata \\
UHS DR3 459785050391 & 22:36:09.71 & +38:14:01.6 & New Discoveries & T8 & 1.6 & \nodata & \nodata \\
UHS DR3 459856516289 & 14:27:38.14 & +31:57:54.3 & New Discoveries & T8 & 1.8 & \nodata & \nodata \\
UHS DR3 459900436213 & 17:08:30.09 & +44:24:57.3 & New Discoveries & T8 & 1.1 & \nodata & \nodata \\
UHS DR3 459985723337 & 02:28:18.01 & +45:25:49.8 & New Discoveries & T8 & 1.5 & \nodata & \nodata \\
CWISE J044853.70-193543.6 & 04:48:53.63 & -19:35:44.5 & New Discoveries & T7 & 1.5 & \nodata & \nodata \\
CWISE J081921.15+740017.4 & 08:19:21.17 & +74:00:17.5 & New Discoveries & T7 & 1.4 & \nodata & \nodata \\
UGCS J122914.34+221448.6 & 12:29:14.34 & +22:14:48.6 & New Discoveries & T7 & 1.5 & \nodata & \nodata \\
CWISE J012055.42+752710.5 & 01:20:55.43 & +75:27:10.5 & New Discoveries & T7 & 1.4 & \nodata & \nodata \\
CWISE J111858.30-242304.1 & 11:18:58.31 & -24:23:04.1 & New Discoveries & T7 & 1.9 & \nodata & \nodata \\
CWISE J131932.56-075412.6 & 13:19:32.56 & -07:54:12.7 & New Discoveries & T7 & 1.6 & \nodata & \nodata \\
CWISE J220948.71+235913.2 & 22:09:48.71 & +23:59:13.2 & New Discoveries & T7 & 1.5 & \nodata & \nodata \\
UHS DR3 459778426590 & 20:37:46.99 & +14:23:57.3 & New Discoveries & T7 pec & 4.7 & \nodata & \nodata \\
UHS DR3 459800102948 & 12:22:42.76 & +53:34:15.2 & New Discoveries & T7 & 1.2 & \nodata & \nodata \\
CWISE J060305.29+543836.8 & 06:03:05.30 & +54:38:36.9 & New Discoveries & T6 & 1.2 & \nodata & \nodata \\
CWISE J165709.68+261334.6 & 16:57:09.68 & +26:13:34.7 & New Discoveries & T6 & 1.3 & \nodata & \nodata \\
UHS DR3 459597400384 & 20:06:50.46 & +12:52:21.2 & New Discoveries & T6 & 1.2 & \nodata & \nodata \\
UHS DR3 459661307265 & 10:07:50.45 & +47:14:15.0 & New Discoveries & T6 & 1.4 & \nodata & \nodata \\
UHS DR3 459686287978 & 06:08:52.54 & +46:07:53.0 & New Discoveries & T6 & 1.4 & \nodata & \nodata \\
UHS DR3 459765697995 & 23:39:17.64 & +50:41:15.8 & New Discoveries & T6 & 1.8 & \nodata & \nodata \\
UHS DR3 459988753959 & 20:51:52.03 & +10:38:06.9 & New Discoveries & T6 & 1.4 & \nodata & \nodata \\
CWISE J022243.38+323137.2 & 02:22:43.39 & +32:31:37.4 & New Discoveries & T5 & 1.9 & \nodata & \nodata \\
\enddata
\tablecomments{References: (1) \cite{2018ApJS..236...28T}; (2) This paper; (3) \cite{2015MNRAS.450.2486C}; (4) \cite{2022arXiv220800070G}; (5) \cite{2024ApJS..271...55K}; (6) \cite{2012ApJ...753..156K}; (7) \cite{2012ApJ...759...60T}; (8) \cite{2013ApJS..205....6M}; (9) \cite{2018ApJ...867..109M}; (10) \cite{2011ApJS..197...19K}; (11) \cite{2009MNRAS.397..258L}; (12) \cite{2025ApJ...982...79B}; (13) \cite{2014AJ....147..113C}; (14) \cite{2013AJ....145...84W}; (15) \cite{2008MNRAS.391..320B}; (16) \cite{2003ApJ...594..510B}; (17) \cite{2011ApJ...735..116B}; (18) \cite{2020ApJ...899..123M}; (19) \cite{2021ApJS..253....7K}; (20) \cite{2023ApJ...958...94R}; (21) \cite{2020ApJ...889...74M}; (22) \cite{2011AJ....141..203A}; (23) \cite{2013MNRAS.433..457B}; (24) \cite{2007AJ....134.1162L}; (25) \cite{2015ApJ...804...92S}; (26) \cite{2008MNRAS.390..304P}; (27) \cite{2007MNRAS.379.1423L}; (28) \cite{2013PASP..125..809T}; (29) \cite{2012AA...548A..53L}; (30) \cite{2006ApJ...637.1067B}; (31) \cite{2010MNRAS.406.1885B}; (32) \cite{2015ApJ...814..118B}; (33) \cite{2006AJ....131.2722C}; (34) \cite{2016ApJ...826...73P}; (35) \cite{2019AJ....158..182G}; (36) \cite{2012ApJ...757..100D}; (37) \cite{2010ApJ...710.1142B}; (38) \cite{2014AJ....148..129A}; (39) \cite{2017ApJ...841L..19K}; (40) \cite{2013AA...557A..43B}; (41) \cite{2010ApJ...718L..38A}; (42) \cite{2007ApJ...663..677S}; (43) \cite{2010AA...522A.112R}; (44) \cite{2011AJ....142...77D}; (45) \cite{2019MNRAS.486.1260Z}; (46) \cite{2025MNRAS.542..656Z}; (47) \cite{2024ApJ...967..147M}; (48) \cite{2022AJ....163..242S}; (49) \cite{2022ApJ...924...68V}; (50) \cite{2017AJ....154..112K}; (51) \cite{2009AJ....137..304S}; (52) \cite{2015MNRAS.449.3651M}; (53) \cite{2011ApJ...743...50C}; (54) \cite{2007AA...466.1059K}; (55) \cite{2013AJ....146..161M}; (56) \cite{2013ApJ...777...84B}; (57) \cite{2013MNRAS.430.1171D}; (58) \cite{2023ApJ...959...63S}; (59) \cite{2014MNRAS.443.2327S}; (60) \cite{2019ApJ...883..183M}; (61) \cite{2016ApJ...833...96L}; (62) \cite{2013ApJ...778...36R}; (63) \cite{2008ApJ...689.1295K}; (64) \cite{2008ApJ...676.1281M}; (65) \cite{2011ApJ...739...48A}; (66) \cite{2014ApJ...787..126L}; (67) \cite{2016ApJ...817..112S}; (68) \cite{2024AA...686A.171Z}; (69) \cite{2016ApJ...830..144R}; (70) \cite{2018PASJ...70S..35M}; (71) \cite{2006ApJ...651L..57A}; (72) \cite{2010ApJS..190..100K}; (73) \cite{2023arXiv230711882S}; (74) \cite{2015MNRAS.454.4054B}; (75) \cite{2020AJ....159..257B}; (76) \cite{2024AJ....168..165S}; (77) \cite{2002AJ....123.3409H}; (78) \cite{2008ApJ...685.1183L}; (79) \cite{2012ApJS..201...19D}; (80) \cite{2015AJ....150..182K}; (81) \cite{2008MNRAS.385L..53C}; (82) \cite{2004AA...416L..17K}; (83) \cite{2018MNRAS.480.5447Z}; (84) \cite{2015ApJS..219...33G}; (85) \cite{2023AJ....166..226B}; (86) \cite{1999ApJ...519..802K}; (87) \cite{2014AJ....147...34S}; (88) \cite{2014AA...567A..43S}; (89) \cite{2014ApJ...783..122K}; (90) \cite{2007AJ....133..439C}; (91) \cite{2000AJ....120..447K}; (92) \cite{2003AJ....126.2421C}; (93) \cite{2012AJ....144..180M}; (94) \cite{2004AJ....127.3553K}; (95) \cite{2013ApJ...776..126C}; (96) \cite{2014ApJ...792..119D}; (97) \cite{2017AJ....153..196S}; (98) \cite{2013ApJ...772..129B}; (99) \cite{2019MNRAS.484.5142P}; (100) \cite{2009AA...494..949S}; (101) \cite{2009AA...497..619Z}; (102) \cite{2008ApJ...674..451B}; (103) \cite{2016ApJS..224...36K}; (104) \cite{2011ApJ...732...56G}; (105) \cite{2015AJ....150..179G}; (106) \cite{2024AJ....167..253R}; (107) \cite{2019ApJ...883..205B}; (108) \cite{2014ApJ...787....5N}; (109) \cite{2017ApJ...843L...4B}; (110) \cite{2017ApJS..228...18G}; (111) \cite{2012AA...539A.151A}; (112) \cite{2016ApJS..225...10F}; (113) \cite{2013ApJ...777L..20L}; (114) \cite{2023ApJ...943L..16S}; (115) \cite{2008AJ....136.1290R}; (116) \cite{2018AA...620A.130P}; (117) \cite{2018RNAAS...2..205M}; (118) \cite{2014ApJ...794..143B}; (119) \cite{2010AJ....139.1808S}; (120) \cite{2021MNRAS.506.1944B}; (121) \cite{2019AJ....157..231K}; (122) \cite{2017AA...599A..78P}; (123) \cite{2009AJ....137.3345C}; (124) \cite{2013AJ....145....2F}; (125) \cite{2008ApJ...686..528L}; (126) \cite{2022ApJ...935...15Z}; (127) \cite{2013MNRAS.431.2745G}; (128) \cite{2016ApJ...816...78C}; (129) \cite{2024AJ....168..159L}; (130) \cite{2018AJ....155...34C}; (131) \cite{2023AJ....165...37L}; (132) \cite{2003AA...403..929K}; (133) \cite{2010AJ....139..176F}; (134) \cite{2008MNRAS.383..831P}; (135) \cite{2022ApJ...926L..12S}; (136) \cite{2016ApJ...822L...1S}; (137) \cite{2025AJ....170...19L}; (138) \cite{2024ApJ...961..121H}; (139) \cite{2016ApJ...821..120A}; (140) \cite{2018AJ....156...76L}; (141) \cite{2010AA...519A..93B}; (142) \cite{2017AJ....154...46E}; (143) \cite{2001AJ....121.3235K}; (144) \cite{2009AJ....137....1F}; (145) \cite{2014MNRAS.445.3908L}; (146) \cite{2000AJ....119..928F}; (147) \cite{2003PASP..115.1207T}; (148) \cite{2022AA...664A.111B}; (149) \cite{2002ApJ...575..484G}; (150) \cite{2009AIPC.1094..561S}; (151) \cite{2009AJ....138.1563B}; (152) \cite{2018RNAAS...2...50C}; (153) \cite{2016PhDT.......189A}; (154) \cite{2010MNRAS.404.1817Z}; (155) \cite{2006ApJ...645..676L}; (156) \cite{2020AA...640A...9M}; (157) \cite{2013MNRAS.435.2650C}; (158) \cite{2011AA...527A..24L}; (159) \cite{2020AA...633A.152C}; (160) \cite{2013AA...549A.123A}; (161) \cite{2000ApJ...543..299M}; (162) \cite{2000Sci...290..103Z}; (163) \cite{2012MNRAS.427.3280F}; (164) \cite{2010AA...517A..53M}; (165) \cite{2015ApJ...802...37B}; (166) \cite{2002ApJ...564..466G}; (167) \cite{2008AJ....135..785W}; (168) \cite{2017ApJ...837...95B}; (169) \cite{2007MNRAS.374..445K}; (170) \cite{2020MNRAS.494.4891M}; (171) \cite{2024AA...685A...6R}; (172) \cite{2017AJ....153...92T}; (173) \cite{2022AA...657A.129A}; (174) \cite{2021MNRAS.503.2265L}; (175) \cite{2016ApJ...820...32B}; (176) \cite{2003ApJ...586L.149S}; (177) \cite{1999AJ....118.2466M}; (178) \cite{2019AJ....157..234S}; (179) \cite{2012ApJ...752...56F}; (180) \cite{2005AA...440.1061L}; (181) \cite{2003IAUS..211..197W}; (182) \cite{2002AJ....123..458S}; (183) \cite{2007ApJ...658..557B}; (184) \cite{2000AJ....120.1085G}; (185) \cite{2012ApJ...760..152L}; (186) \cite{2021MNRAS.507.4646K}; (187) \cite{2020ApJS..249....3A}; (188) \cite{2024MNRAS.534..695C}; (189) \cite{2015AJ....149..158S}; (190) \cite{2019CoSka..49..546C}; (191) \cite{2016AA...589A..49S}; (192) \cite{2018AA...619L...8R}; (193) \cite{2017AJ....153...46L}; (194) \cite{2018ApJ...858...41Z}; (195) \cite{2014MNRAS.442.1586D_DUP}; (196) \cite{2020AJ....160...44L}; (197) \cite{2020PASP..132j4401A}; (198) \cite{2017MNRAS.464.3040Z}; (199) \cite{2019MNRAS.489.1423Z}; (200) \cite{2011AJ....141...97W}; (201) \cite{2004AJ....127.2856B}; (202) \cite{1995AJ....109..797K}; (203) \cite{2004ApJ...600.1020M}; (204) \cite{1991ApJ...367L..59R}; (205) \cite{2006MNRAS.366L..40P}; (206) \cite{2013ApJ...779..172G}; (207) \cite{2004AA...425..519S}; (208) \cite{1997ApJ...476..311K}; (209) \cite{2013AA...556A..15R}; (210) \cite{2022ApJ...941..101S}; This table is available in its entirety online; the full table contains 16917 rows.}
\end{deluxetable*}